\documentclass[prb,preprint,superscriptaddress]{revtex4-1} 

\usepackage{amsmath}  
\usepackage{amsfonts} 
\usepackage{graphicx} 
\usepackage{subfigure}
\usepackage{float}
\usepackage{dcolumn}
\usepackage{CJKutf8}      
\usepackage[colorlinks,linkcolor=blue]{hyperref}

\begin{document}


\title{Nonlinear effect of forced harmonic oscillator subject to sliding friction and simulation by a simple nonlinear circuit}

\author{Qian Xu}
\affiliation{School of Physics, Nanjing University, Nanjing, P.R China, 210093}
\author{Wenkai Fan}
\altaffiliation[Current address: ]{Physics Bldg, Durham, NC 27710}
\affiliation{Kuangyanming Honor School, Nanjing University, Nanjing, P.R China, 210093}
\email[]{wf39@duke.edu}

\author{Sihui Wang}
\email[]{wangsihui@nju.edu.cn}
\affiliation{School of Physics, Nanjing University, Nanjing, P.R China, 210093}

\author{Hongjian Jiang}
\affiliation{School of Physics, Nanjing University, Nanjing, P.R China, 210093}


\date{\today}

\begin{abstract}
We study the nonlinear behaviors of mass-spring systems damped by dry friction using simulation by a nonlinear LC circuit damped by anti-parallel diodes. We show that the differential equation for the electric oscillator is equivalent to the mechanical one's when a piecewise linear model is used to simplify the diodes I-V curve. We derive series solutions to the differential equation under weak nonlinear approximation which can describe the resonant response as well as amplitudes of superharmonic components. Experimental results are consistent with series solutions. We also present the phenomenon of hysteresis. Theoretical analysis along with numerical simulations are conducted to explore the stick-slip boundary. 

The correspondence between the mechanical and electric oscillators makes it easy to demonstrate the behaviors of this nonlinear oscillator on a digital oscilloscope. It can be used to extend the linear RLC experiment at the undergraduate level.
\end{abstract}

\maketitle 

\section{Introduction} 

The phenomenon of dry friction is common in fundamental physics courses as well as in the real world. However, when combined with an ordinary mass-spring system, nonlinearity arises due to discontinuity of dry friction at zero velocity. The behaviors of free oscillation of mass-spring systems damped by dry friction have been studied in many previous papers.\cite{ref1,ref2,ref3,ref4,ref5,ref6,ref7,ref8,ref9,ref10,ref11} Among them, Jiang\cite{ref11} describes the correspondence between a mechanical oscillator and an electric one by constructing a varied LC circuit damped by anti-parallel diodes. The mechanical and electric oscillators’ transient oscillation curves and phase trajectories are similar to each other. In these works, however, nonlinearity is not emphasized.

On the other hand, over decades, numerous works have been devoted to introducing nonlinear phenomena to students and scientists.\cite{ref13} Some introduce analytical methods to solve nonlinear problems \cite{ref14,ref15} and describe different experimental demonstrations for undergraduates.\cite{ref16,ref17,ref18,ref19,ref20} Usually, nonlinearity comes from the restoring force containing a cubic term (Duffing’s type)\cite{ref16,ref17} or piecewise linear terms.\cite{ref18} These papers emphasize the amplitude frequency response together with the “jump” or hysteresis phenomenon. The oscillation with nonlinear damping term depending on velocity is occasionally considered.\cite{ref20} In addition, the transient and steady behavior of a nonlinear damped electric oscillator are studied to show amplitude dependence.\cite{ref12}

The purpose of this paper is to study the nonlinear behaviors of a mass-spring oscillator damped by sliding friction. We use an LC circuit damped by anti-parallel diodes to simulate the mechanical model. Using the piecewise linear model to simplify the diodes I-V curve, we find that the differential equation for the electric oscillator is equivalent to that of the mechanical one. We obtain the series solution to the differential equation and describe the resonant response and the amplitudes of superharmonic components. Our series solutions are  consistent with experimental results. The phenomenon of hysteresis is also presented. 

The experiment described in this paper comprises of part of the fundamental test problems in the $4^{th}$ National College Students’ Physics Experimental Competition in China. The mechanical model and the analytical solution we present are conceptually simple. The correspondence between the mechanical model and the simple electric circuit makes it easy to observe the nonlinear oscillation curves and the frequency spectrum of harmonics on a digital oscilloscope. It can be used to extend the linear RLC experiment at the undergraduate level.

\section{the theoretical model}

\begin{figure}[H]
\centering
\subfigure[]{
\label{fig:f-v_curve:a}
\includegraphics[width = 0.3\textwidth]{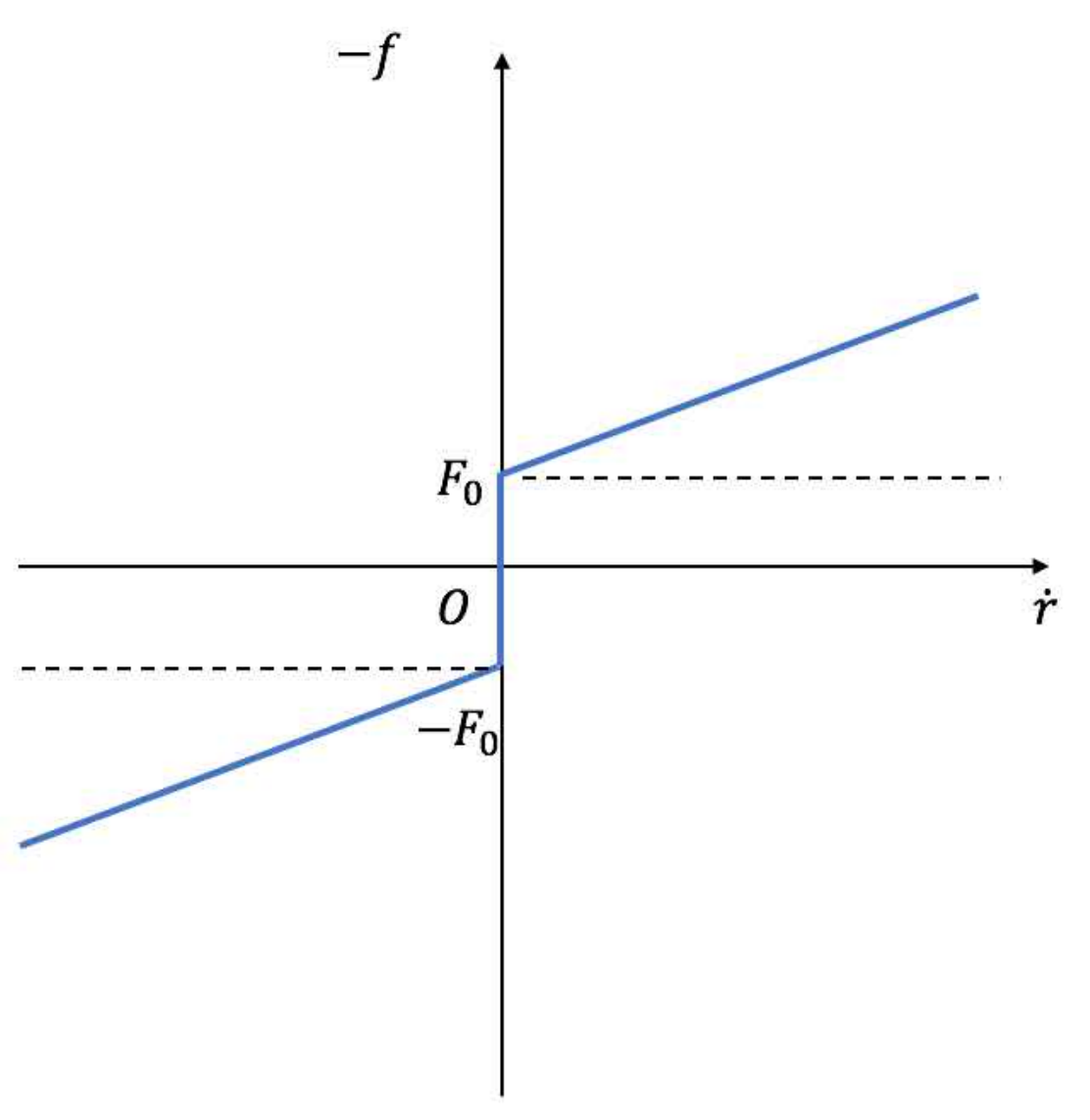}}
\subfigure[]{
\label{fig:f-v_curve:b}
\includegraphics[width = 0.3\textwidth]{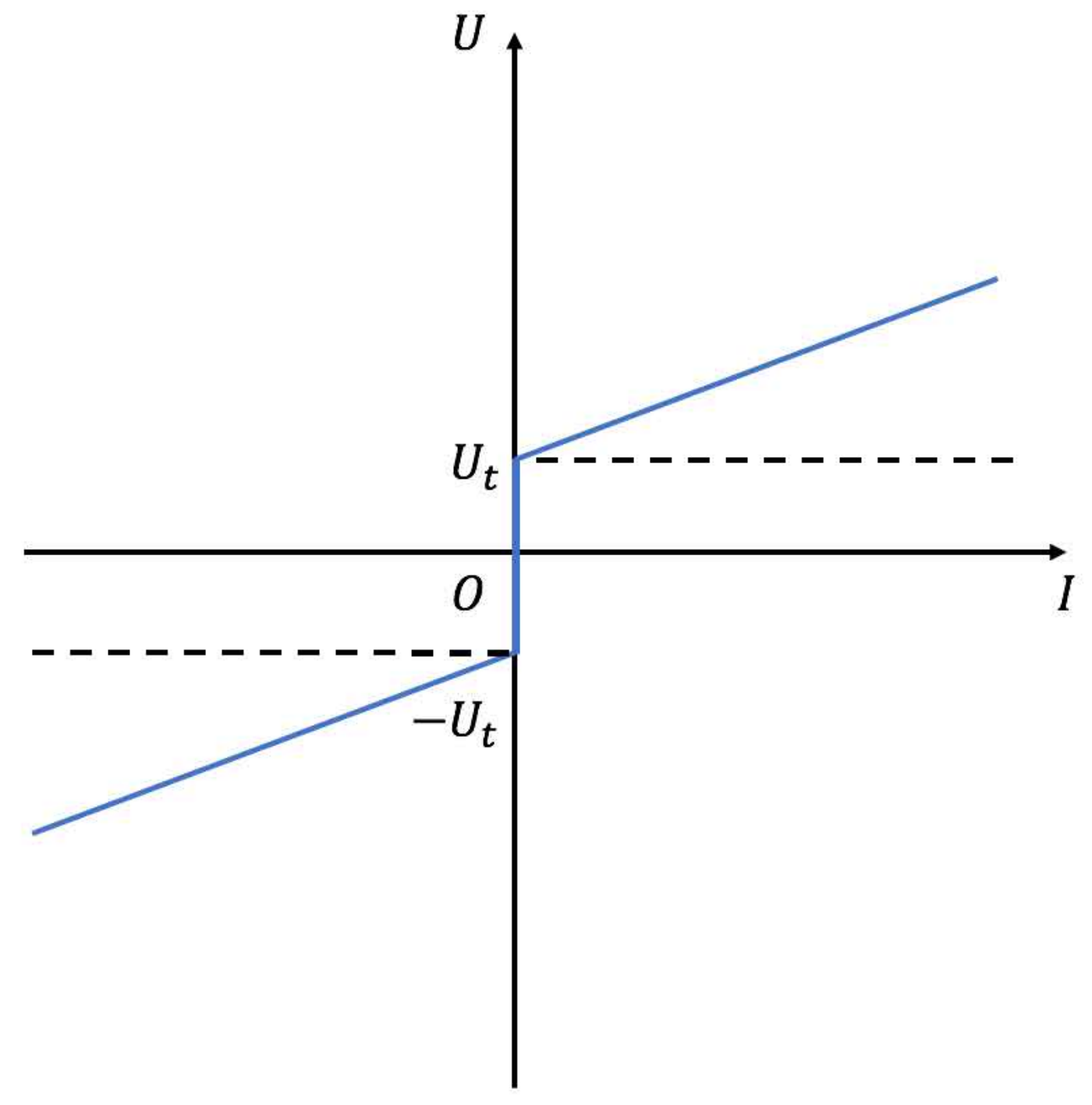}}
\caption{(a) Velocity dependent friction force (b) Voltage-current curve of the anti-parallel diodes using PWL model}
\label{fig:f-v_curve}
\end{figure}

The equation of motion of a forced oscillator subject to dry friction is
\begin{equation}
m\ddot{r} + kr  = F{\rm cos}(\Omega t + \phi_0) + f(\dot{r})
\label{eq:dynamical raw model}
\end{equation}
with 
\begin{equation}
    f(\dot{r}) = \left\{
    \begin{aligned}
        &-F_0 - \gamma \dot{r},\ \ \ \dot{r} > 0\\
        &f_0,\ \ \ \ \ \ \ \ \ \ \ \ \ \ \dot{r} = 0\\
        & F_0 - \gamma \dot{r},\ \ \ \ \ \ \ \dot{r} < 0\\
    \end{aligned}
    \right.
\end{equation}
where $m$ is the mass of the oscillator, $r$ is the displacement, $k$ is the spring constant, $f(\dot{r})$ is the velocity dependent friction force, $F_0$ is the magnitude of sliding friction force, and $f_0$ is the static friction which satisfies $|f_0| < F_0$, see Fig.~\ref{fig:f-v_curve:a}. Here we assume that the static and kinetic friction coefficients are the same. We also include a linear damping term $\gamma \dot{r}$, where $\gamma$ is the damping coefficient. $F$, $\Omega$, $\phi_0$ are the amplitude, the angular frequency, and the initial phase of the periodical driving force, respectively. 

When only considering the sliding motion, Eq.~\ref{eq:dynamical raw model} can be simplified to:
\begin{equation}
m\ddot{r} + kr + F_0 {\rm sgn}(\dot{r}) + \gamma \dot{r} = F{\rm cos}(\Omega t + \phi_0) 
\label{eq:dynamical model}
\end{equation}

\begin{figure}[H]
    \centering
    \subfigure[]{
    \label{fig:circuit_diagram:schematica diagram}
    \includegraphics[width = 0.35\textwidth,height = 0.3\textwidth]{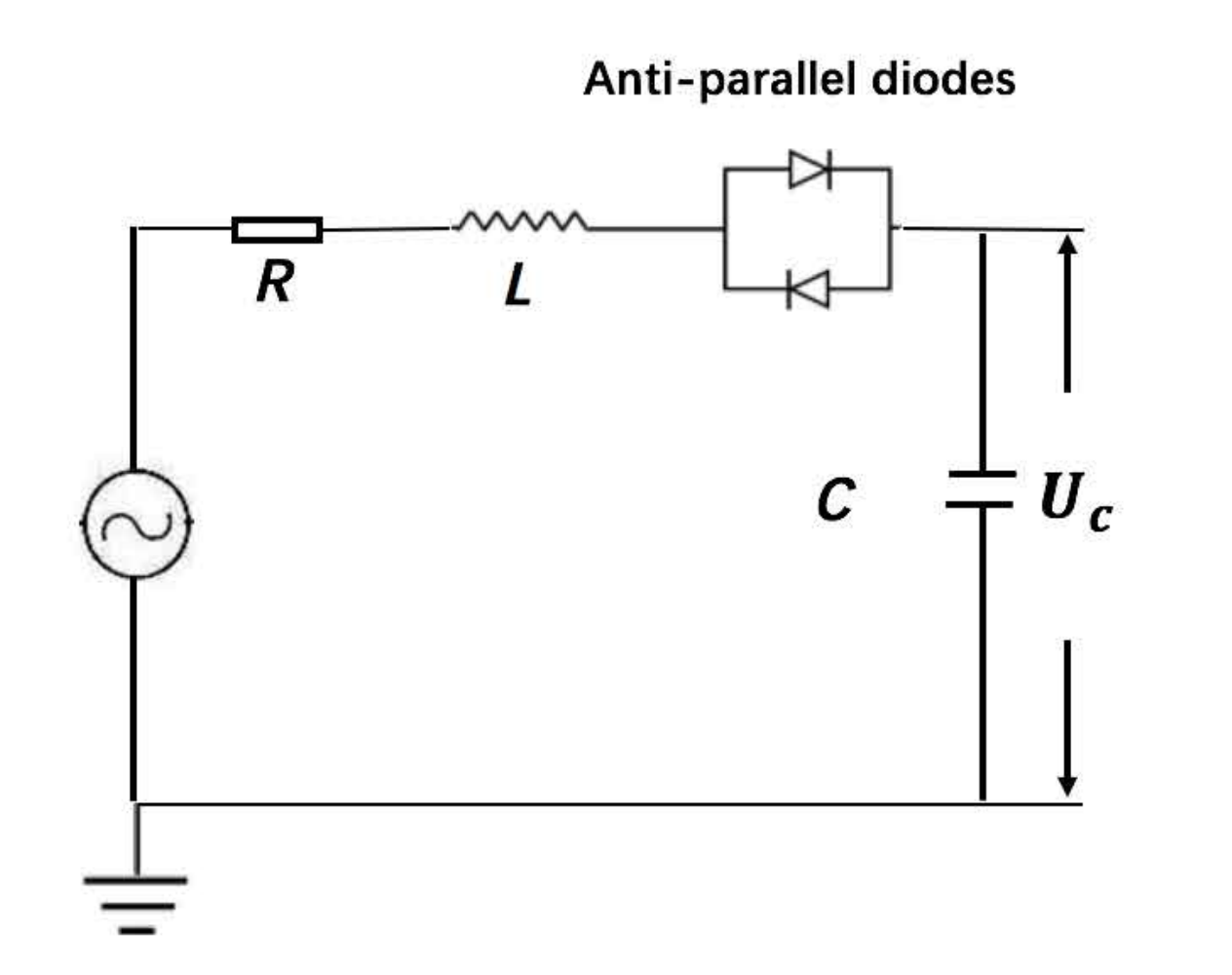}}
    \subfigure[]{
    \label{fig:circuit_diagram:PWL}
    \includegraphics[width = 0.5\textwidth,height = 0.25\textwidth]{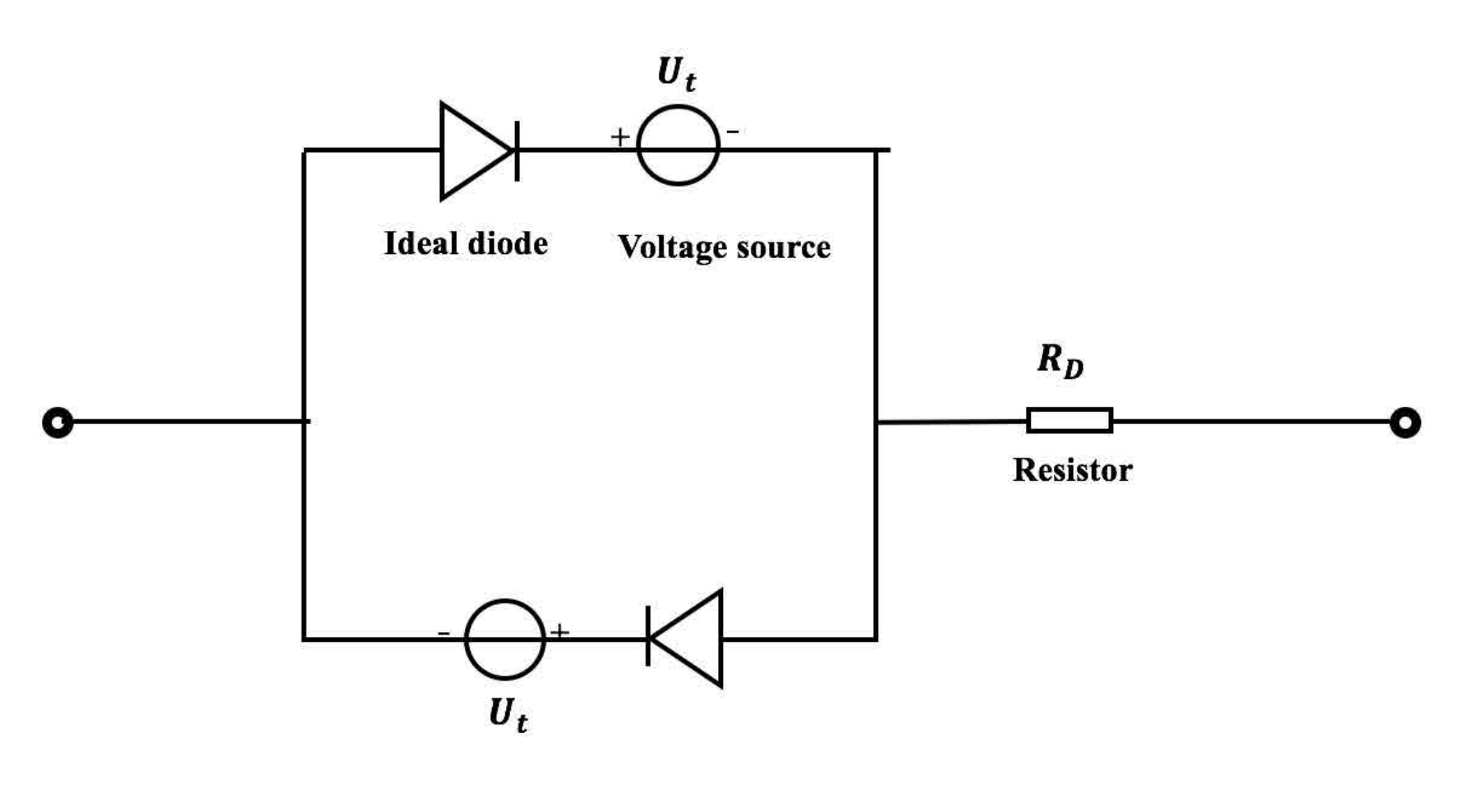}}
    \caption{(a) RLC circuit with anti-parallel diodes (b) PWL model of anti-parallel diodes}
    \label{fig:circuit_diagram}
\end{figure}

Fig.~\ref{fig:circuit_diagram:schematica diagram} illustrates an electric circuit analogous to this mechanical oscillator, in which a pair of anti-parallel diodes is added to the standard RLC circuit.

We use the piecewise linear (PWL)\cite{wiki} model  to simplify the I-V characteristic curve of the diode. In the PWL model, a real diode is modelled as three components in series: an ideal diode, a voltage source, and a resistor.\cite{wiki} Thus the circuit in Fig.~\ref{fig:circuit_diagram:PWL}) is equivalent to the anti-parallel diodes. Hence, the voltage-current curve for anti-parallel diodes has exactly the same form as the friction-velocity curve of the mechanical oscillator (see Fig.~\ref{fig:f-v_curve:a},~\ref{fig:f-v_curve:b}). The differential equation of the circuit in Fig.~\ref{fig:circuit_diagram:schematica diagram} is:
\begin{equation}
\centering
    L\ddot{q} + \frac{1}{C}q + U_t {\rm sgn}(\dot{q}) + (R+R_D)\dot{q} = U{\rm cos}(\Omega t + \phi_0)
\label{eq:electrical model}
\end{equation}
 where $L$ is the inductance, $q$ is the capacitor’s charge, $C$ is the capacitance, $R$ is the linear resistance of the circuit, $U_t$ and $R_D$ are diodes’ turn-on voltage and resistance in PWL model, and $U, \Omega, \phi_0$ are the amplitude, the angular frequency and the initial phase of the sinusoidal voltage source.

Apparently, Eq.~\ref{eq:electrical model} has exactly the same form as the mechanical equation Eq.~\ref{eq:dynamical model} when mechanical variables are replaced by corresponding electrical ones. The correspondences are: capacitor’s charge $q$ for displacement $r$; current  for velocity; inductance $L$ for mass $m$; inverse of capacitance $\frac{1}{C}$ for spring constant $k$; driving voltage $U$ for driving force $F$; diode’s turn-on voltage $U_t$ for friction force $F_0$. When considering the transient response, the dynamics of the electric oscillator are similar to those of the mechanical one\cite{ref11}. Therefore, the PWL model is reasonable and the electric analogy is valid. 

Eq.~\ref{eq:dynamical model} and Eq.~\ref{eq:electrical model} can be both rewritten as:
\begin{equation}
\centering
\ddot{x} + \omega_0^2 x + \alpha \dot{x} +  {\rm sgn}(\dot{x}) = \beta {\rm cos}(\Omega t + \phi_0)
\label{eq:dimensionless eq}
\end{equation}

The correspondence between variables in Eq.~\ref{eq:dimensionless eq} and variables in Eq.~\ref{eq:dynamical model}, Eq.~\ref{eq:electrical model} is shown in Table~\ref{tab:correspondence}. The gain $G$ is the ratio between the capacitor’s voltage amplitude and the source voltage ($\frac{U_{c_0}}{U}$) in the circuit, or $\frac{\omega_0^2r_{0}}{F/m}$ in the mechanical model. $U_{c_0}$ denotes the amplitude of the capacitor's voltage $U_c$, and $r_0$ denotes the amplitude of displacement $r$.
\begin{table}[H]
    \centering
    \caption{Correspondence table}
    \begin{ruledtabular}
    \begin{tabular}{llllll}
    Eq.~\ref{eq:dimensionless eq} & $x$ & $\omega_0^2$ & $\alpha$ & $\beta$ & G \\
    \hline
    Eq.~\ref{eq:dynamical model} & $\frac{r}{F_0/m}$ & $\frac{k}{m}$ & $\frac{\gamma}{m}$ & $\frac{F}{F_0}$ & $\frac{\omega_0^2r_0}{F/m}$ \\
    \hline
    Eq.~\ref{eq:electrical model} & $\frac{q}{U_t/L}$ & $\frac{1}{LC}$ & $\frac{R+R_D}{L}$ & $\frac{U}{U_t}$ & $\frac{U_{c_0}}{U}$\\
    \end{tabular}
    \end{ruledtabular}
    \label{tab:correspondence}
\end{table}

If we neglect the nonlinear damping term $sgn(\dot{x})$ in Eq.~\ref{eq:dimensionless eq},  the approximate steady state solution of the linear forced oscillator is 
\begin{equation}
    x = A_1{\rm cos}\Omega t
    \label{eq:approximate solution}
\end{equation}

The initial phase in Eq.~\ref{eq:dimensionless eq} can be chosen so that the initial phase in Eq.~\ref{eq:approximate solution} is zero. Now we consider the weak nonlinear condition when $\beta\gg1$, and substitute Eq.~\ref{eq:approximate solution} into Eq.~\ref{eq:dimensionless eq} to estimate the damping term $sgn(\dot{x})$.  The “velocity” changes signs twice in each period, so does the damping term $sgn(\dot{x})$. Consequently, the “friction” is exerted to the oscillator in the manner of a square wave at driving angular frequency $\Omega$. Since we are more concerned with the nonlinear behaviors, when neglecting the linear damping, we can convert Eq.~\ref{eq:dimensionless eq} to the following form by using Fourier series expansion of the square waves  
\begin{equation}
    \ddot{x} + \omega_0^2 x -  \sum_{n=1}^{\infty}\frac{4}{(2n-1)\pi}{\rm sin}(2n-1)\Omega t = \beta {\rm cos}(\Omega t + \phi_0)  
\label{eq:dimensionless fourier eq}
\end{equation}

Obviously, the solution to Eq.~\ref{eq:dimensionless fourier eq} contains components of odd-integer harmonic angular frequencies $\Omega, 3\Omega... $. Hence, we modify the trial solution to 
\begin{equation}
    x = A_1{\rm cos}\Omega t  + \sum_{n=2}^{\infty}A_{2n-1}{\rm sin}(2n-1)\Omega t
\label{eq:sol with no damping}
\end{equation}

Substitute Eq.~\ref{eq:sol with no damping} into Eq.~\ref{eq:dimensionless fourier eq}, and compare the coefficients of each harmonic component, we have
\begin{equation}
\left\{
\begin{aligned}
   A_1 = & \frac{\sqrt{\beta^2-(\frac{4}{\pi})^2}}{\omega_0^2-\Omega^2}\\
   A_{2n-1} = &\frac{4}{(2n-1)\pi[(2n-1)^2\Omega^2 -\omega_0^2]}\\
   \phi_0 = & {\rm arcsin}\frac{4}{\beta \pi}
  \end{aligned}
\right.
\label{eq:simplified final solution}
\end{equation}

When the linear damping term is included, the solution becomes
\begin{equation}
 x(t) = A_1{\rm cos}(\Omega t+\phi_1) + \sum_{n=2}^{\infty}A_{2n-1}{\rm cos}[(2n-1)\Omega t+\phi_{2n-1}]
 \label{eq:fianl solution x}
\end{equation}
where coefficients of each harmonic component is
\begin{equation}
\left\{
\begin{aligned}
    A_1 =&\frac{-4\alpha \Omega + \sqrt{\pi^2\beta^2Z^2-16(\Omega^2-\omega_0^2)^2}}{\pi Z^2},\ \ \ \ \ \ \ \ \ \ \ \phi_1 = 0\\
    A_{2n-1} = &\frac{\frac{4}{(2n-1)\pi}}{\sqrt{[\omega_0^2-(2n-1)^2\Omega^2]^2+\alpha^2(2n-1)^2\Omega^2}},\ \ \phi_{2n-1} = -\frac{\pi}{2} - {\rm arctan}\frac{\alpha (2n-1)\Omega}{\omega_0^2-(2n-1)^2\Omega^2}\\
    \phi_0 = &{\rm arccos}\frac{(\omega_0^2-\Omega^2)A_1}{\beta}
\end{aligned}
\right.
\label{eq:final solution}
\end{equation}
where
$$Z = \sqrt{(\omega_0^2-\Omega^2)^2+\alpha^2\Omega^2}$$

Eq.~\ref{eq:final solution} shows that the superharmonic amplitudes $A_3$, $A_5$,… are independent of the driving amplitude $\beta$. This is different from a Duffing oscillator, for its harmonic components are highly dependent on the principal oscillation amplitude.\cite{ref22} In our model, nonlinear effects are relatively strong for small amplitudes.  

\section{Results}
We set up the electric circuit illustrated in Fig.~\ref{fig:circuit_diagram:schematica diagram}. We use an inductor $L=11.5mH$ with an intrinsic resistance of $17.7\Omega$. The capacitor $C=0.02\mu F$.  Two identical $1N4148$ silicon diodes are used with $U_t = 0.7V$, $R_D = 3.0\Omega$. The total resistance of the circuit is $71\Omega$. We collect data  on a $UTD7102$ digital oscilloscope. The driving signal for steady response is a sinusoidal function. Based on the above parameters, the corresponding   theoretical parameters are: $\alpha = 6173.9, \omega_0 = 65938.0\ rad/s$,  according to Table~\ref{tab:correspondence}.

\subsection{Harmonic components of the solution}
 
\begin{figure}[h!]
     \centering
     \subfigure{
     \includegraphics[width = 0.3\textwidth,height = 0.18\textwidth]{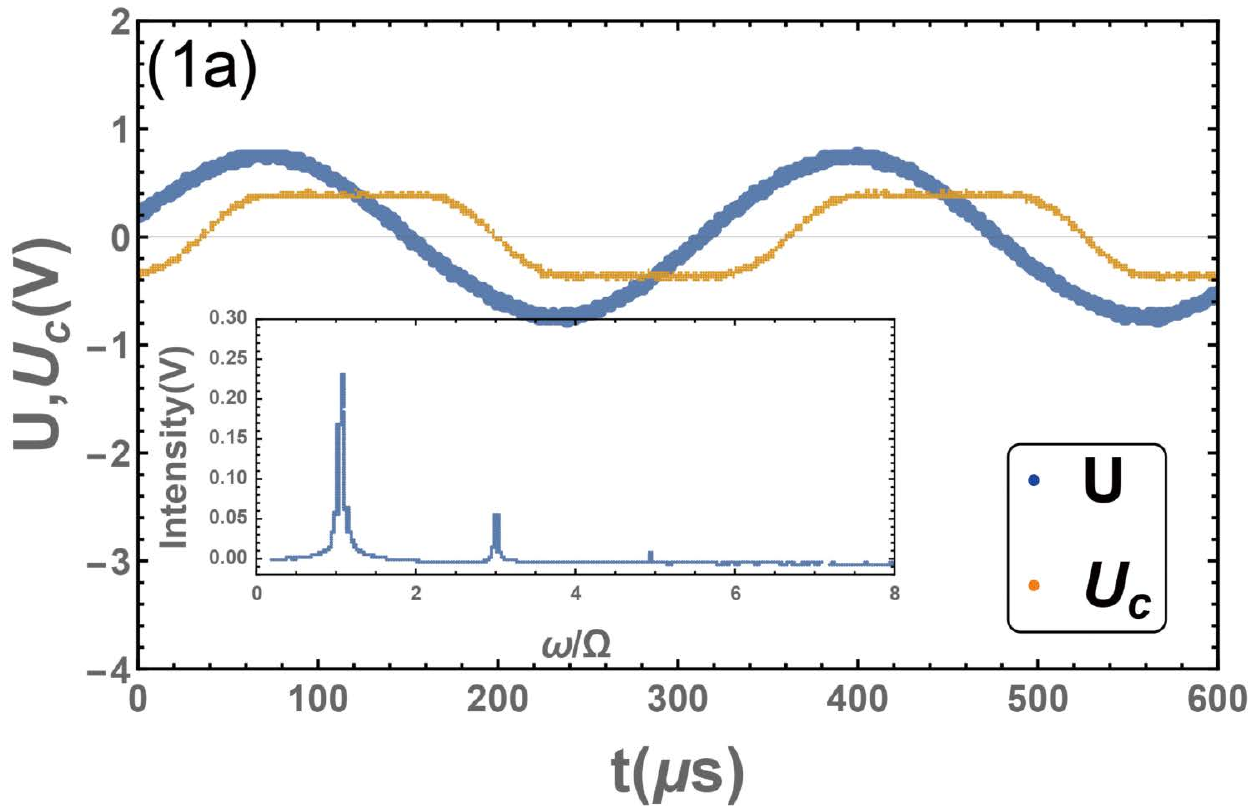}
     \label{1a}
     }
     \subfigure{
     \includegraphics[width = 0.3\textwidth,height = 0.18\textwidth]{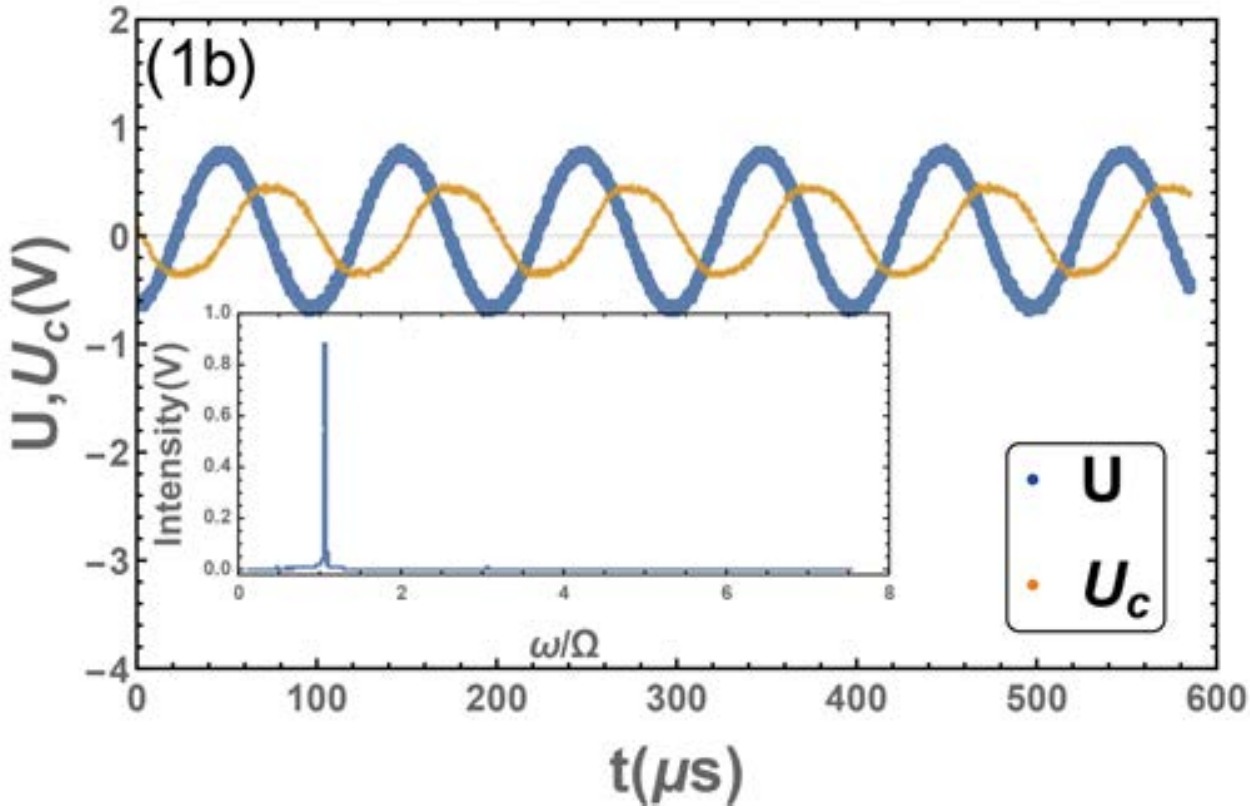}
     \label{1b}
     }
     \subfigure{
     \includegraphics[width = 0.3\textwidth,height = 0.18\textwidth]{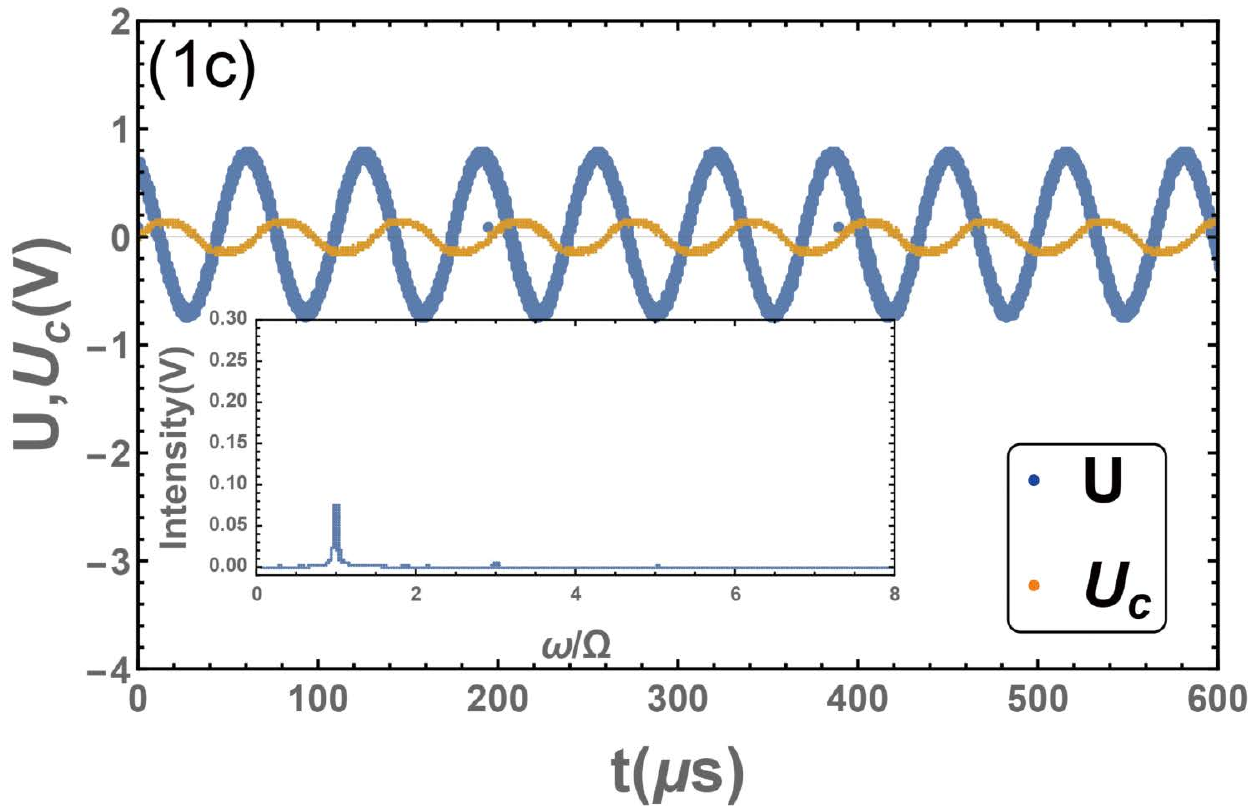}
     \label{1c}
     }
     \hspace{1in}
     \subfigure{
     \includegraphics[width = 0.3\textwidth,height = 0.18\textwidth]{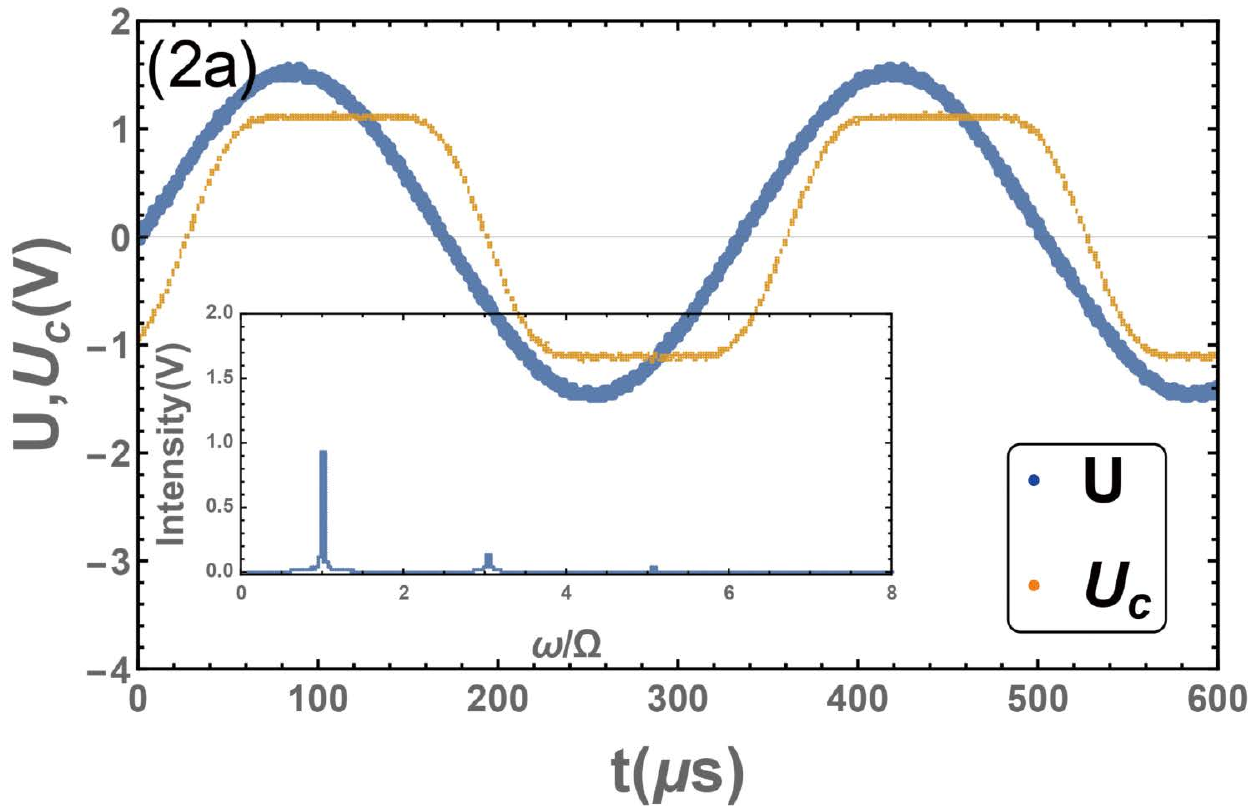}
     \label{2a}
     }
     \subfigure{
     \includegraphics[width = 0.3\textwidth,height = 0.18\textwidth]{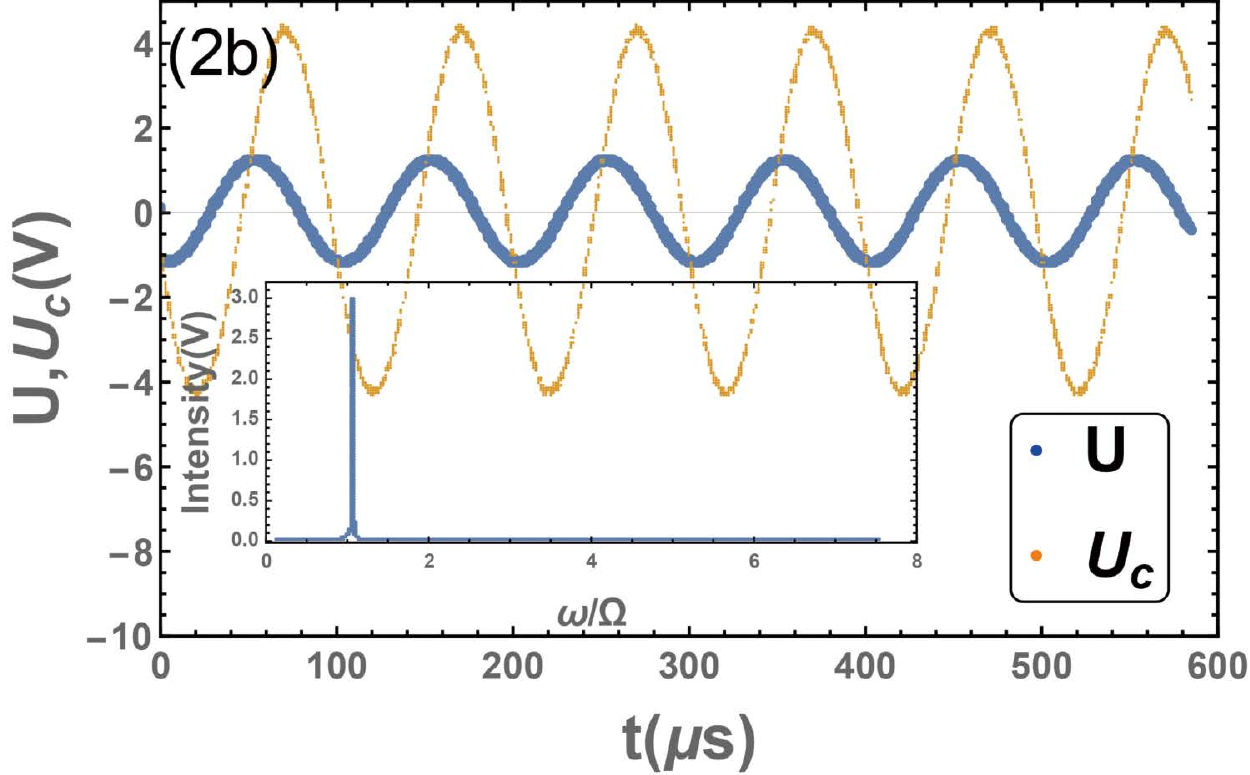}
     \label{2b}
     }
     \subfigure{
     \includegraphics[width = 0.3\textwidth,height = 0.18\textwidth]{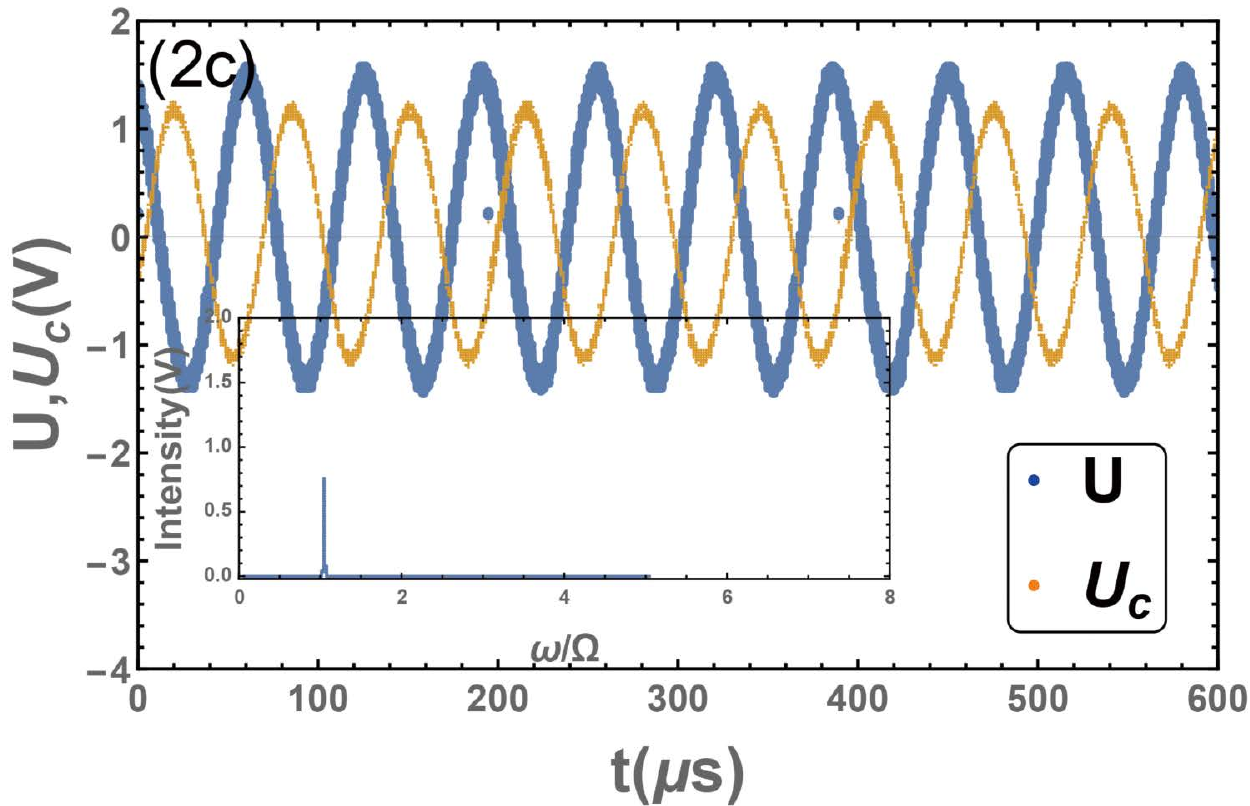}
     \label{2c}
     }
     \hspace{1in}
     \subfigure{
     \includegraphics[width = 0.3\textwidth,height = 0.18\textwidth]{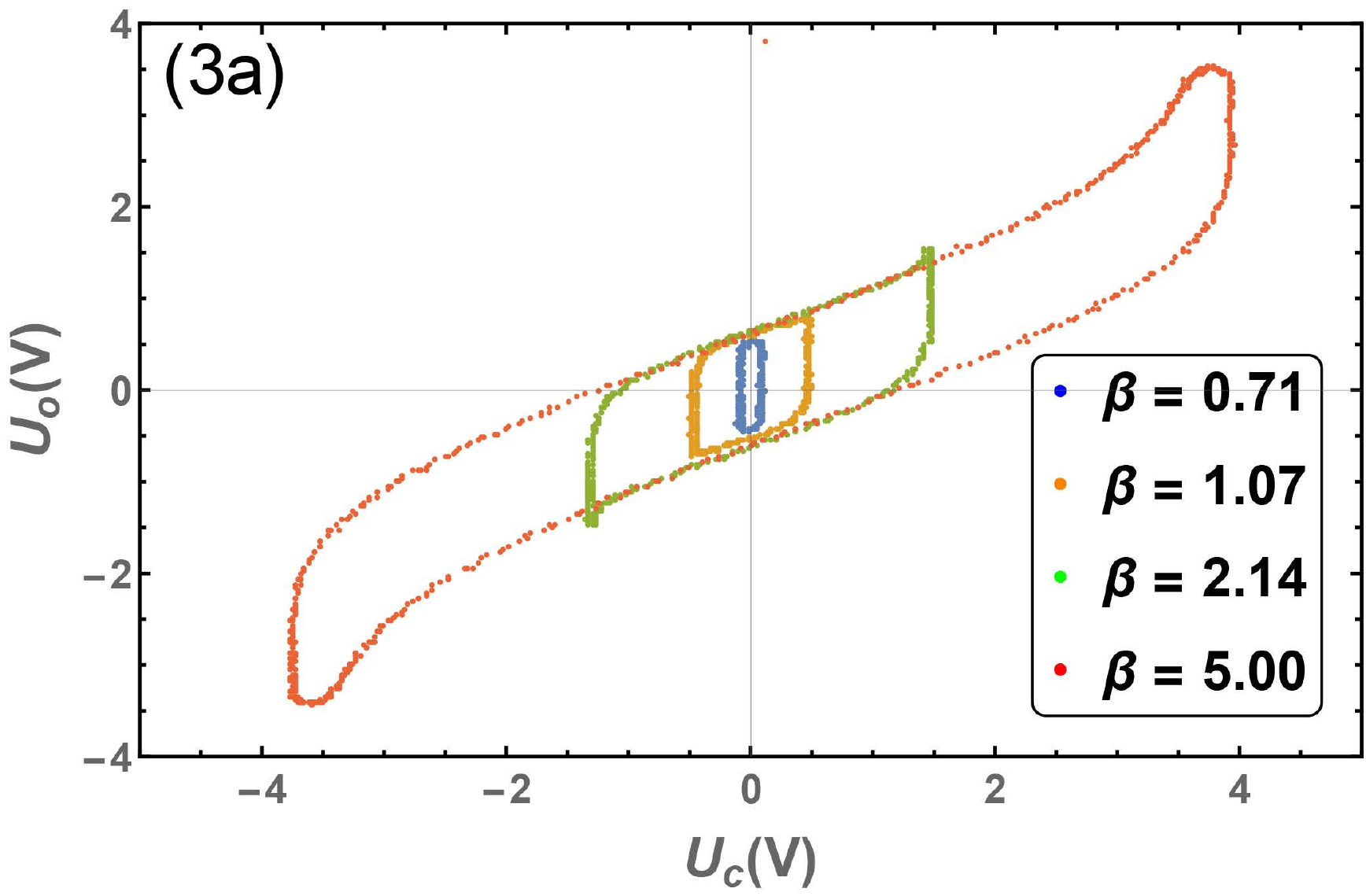}
     \label{2a}
     }
     \subfigure{
     \includegraphics[width = 0.3\textwidth,height = 0.18\textwidth]{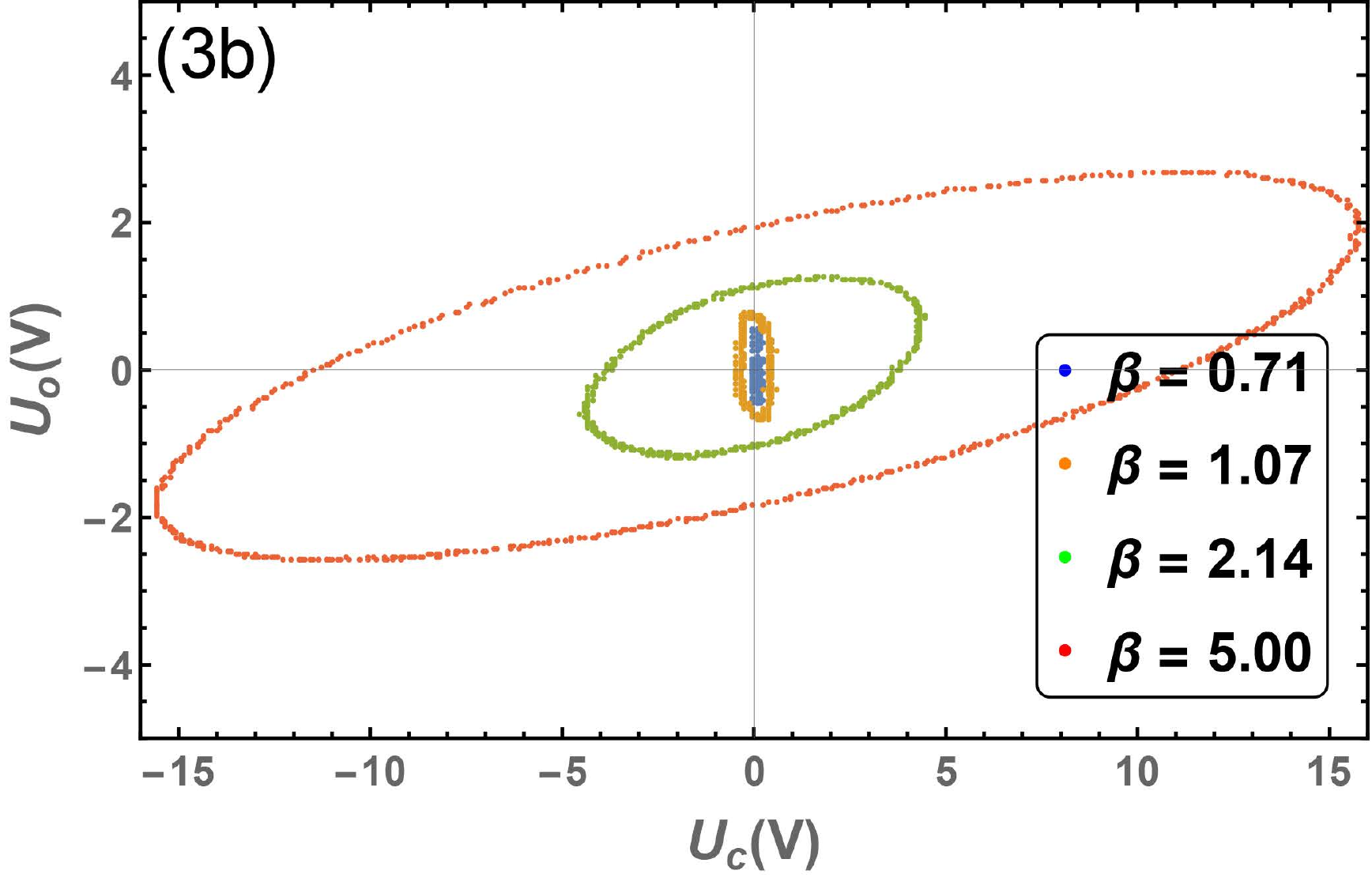}
     \label{2b}
     }
     \subfigure{
     \includegraphics[width = 0.3\textwidth,height = 0.18\textwidth]{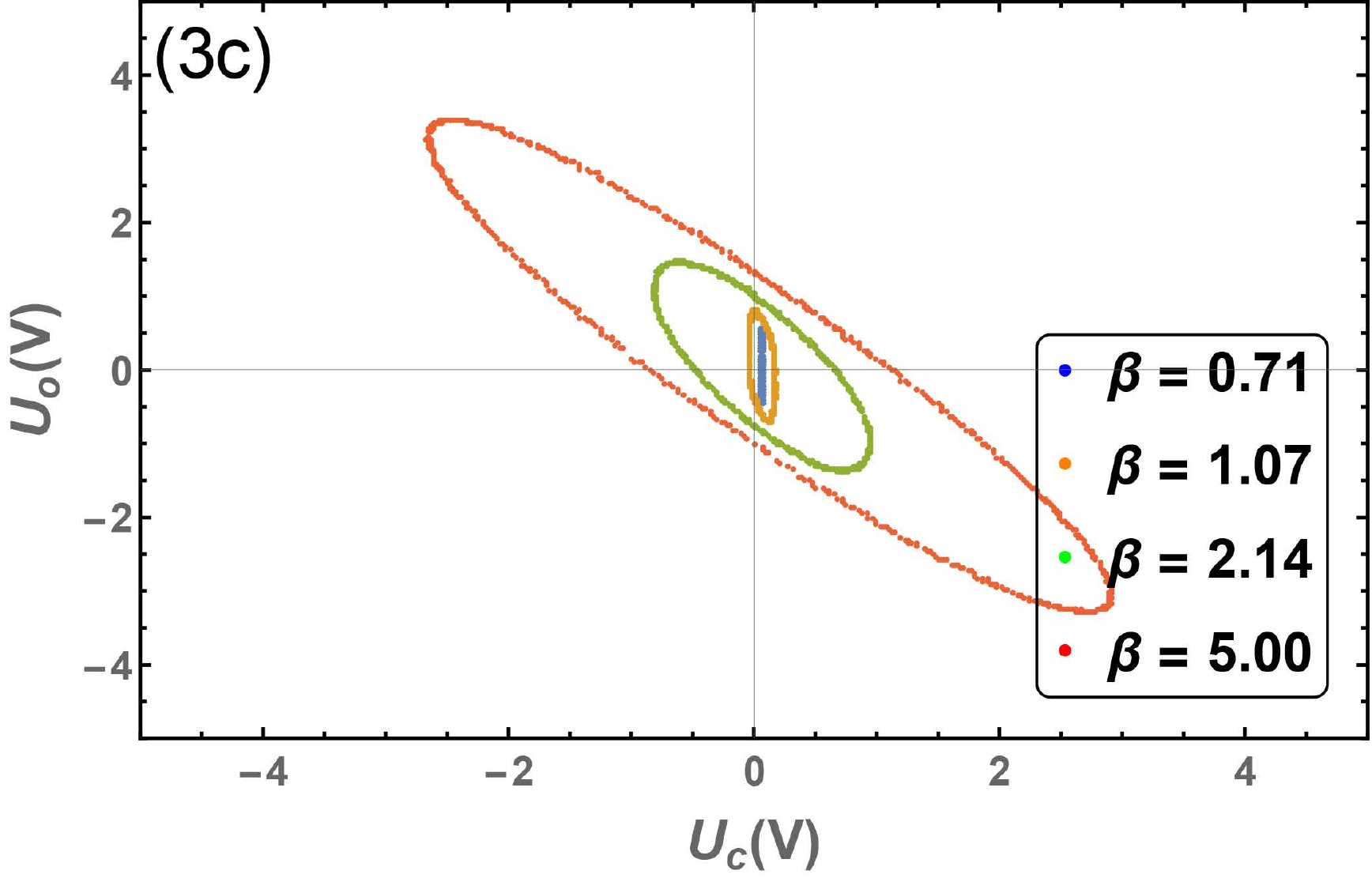}
     \label{2c}
     }
     \caption{(1a)-(1c), (2a)-(2c) are experimental results of the capacitor’s voltage curves and their FFT spectrums under different driving voltage and frequency. (3a)-3(c) are capacitor’s voltage $Uc$ with respect to the driving voltage $U$ in $X-Y$ mode under different driving frequency. 1(a) $\beta = 1.07; \frac{\Omega}{\omega_0} = 0.29$. 1(b) $\beta = 1.07; \frac{\Omega}{\omega_0} = 0.95$. 1(c)
     $\beta = 1.07; \frac{\Omega}{\omega_0} = 1.43$. 2(a) $\beta = 2.14; \frac{\Omega}{\omega_0} = 0.29$. 2(b) $\beta = 2.14; \frac{\Omega}{\omega_0} = 0.95$. 2(c) $\beta = 2.14; \frac{\Omega}{\omega_0} = 1.43$.
     3(a) $\frac{\Omega}{\omega_0} = 0.29$. 3(b) $\frac{\Omega}{\omega_0} = 0.95$. 3(c) $\frac{\Omega}{\omega_0} = 1.43$.}
     \label{fig:group imags}
 \end{figure}

 According to Eqs.~\ref{eq:fianl solution x} -~\ref{eq:final solution}, the solution contains series of harmonic components due to the nonlinearity of the anti-parallel diodes. In the experiment, the capacitors voltage $U_c$, the driving voltage $U$, and the FFT spectrum of $U_c$ are observed on the digital oscilloscope at different driving amplitudes and frequencies. In Figs.~\ref{fig:group imags}(1a)-(1c), the driving amplitude is low ($\beta=1.07$); the driving frequencies are lower than ($\frac{\Omega}{\omega_0}=0.29$), close to ($\frac{\Omega}{\omega_0}=0.95$) and higher than ($\frac{\Omega}{\omega_0}=1.43$) the circuit’s natural frequency, respectively. In Figs.~\ref{fig:group imags}(2a)-(2c), the driving amplitude is higher ($\beta$=2.14), and the driving frequencies are also $\frac{\Omega}{\omega_0}$ = 0.29, 0.95 and 1.43. In Figs.~\ref{fig:group imags}(1)-(2), $U_c$ is denoted by the thin curves and $U$ by the thick curves, and the insets are $U_c$  's FFT spectrums. We see that $U_c$ 's oscillation curves have more distortion at low driving frequency, see Figs.~\ref{fig:group imags}(1a),(2a). In addition, under low frequencies, odd harmonic components at frequencies $\Omega,3\Omega,5\Omega$ are clearly seen in the FFT spectrums. As we have pointed out, unlike other nonlinear oscillators (e.g. Duffing oscillator), the odd harmonic components are relatively stronger for smaller driving amplitudes. Figs.~\ref{fig:group imags}(1b),(2b) show that when the driving frequency is close to the circuit's natural frequency ($\frac{\Omega}{\omega_0}=0.95$), the simple harmonic solution $x = A_1 cos\omega t$ is a good approximation, even when the driving amplitude is very low, see Fig.~\ref{fig:group imags}(1b) when $\beta=1.07$. When the driving frequency is higher than the natural frequency, $\frac{\Omega}{\omega_0}=1.43$, simple harmonic solution is appropriate only when the driving amplitude is high, see Fig.~\ref{fig:group imags}(2c) when $\beta=2.14$. The phase lag between the capacitor’s voltage $U_c$ and the driving voltage $U$ can be viewed on the oscilloscope’s X-Y mode, see Figs.~\ref{fig:group imags}(3a)-(3c). The simple harmonic solution is appropriate when the phase loop forms an ellipse. We discuss hysteresis at low frequency, see Fig.~\ref{fig:group imags}(3a) when $\frac{\Omega}{\omega_0}=0.29$, in Sec.C. 

\subsection{Amplitude-frequency response}
\begin{figure}[h]
\centering
 \includegraphics[width = 0.6\textwidth]{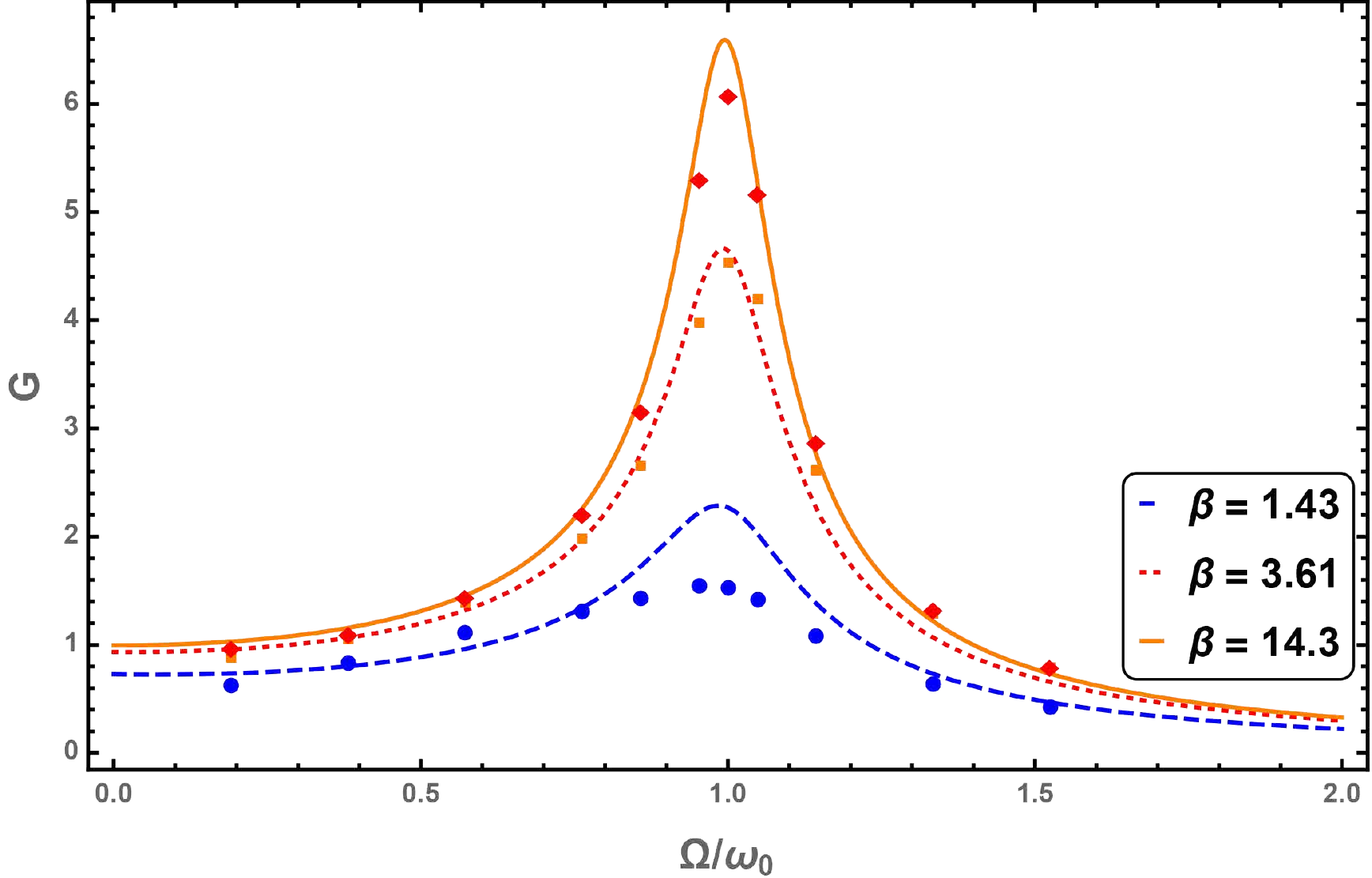}
 \caption{Resonant response of the Gain at driving amplitudes $\beta = 1.43,3.61,14.3$. The data points are experimental results of voltage gain $G=\frac{U_c}{U}$. The curves are theoretical results.}
 \label{fig:resonant response}
 \end{figure}

For linear oscillators, the gain $G$, the sharpness of the resonant peak and the resonant frequency are independent of the driving amplitude. For this nonlinear oscillator, the independence holds only when $\beta\gg1$. When the driving amplitude $\beta$ is reduced, the resonant peak becomes lower and wider, see Fig.~\ref{fig:resonant response}. The peak finally vanishes below a threshold amplitude in the experiment. The data points of the upper, middle and lower cases in Fig.~\ref{fig:resonant response} are the experimental results for $\beta = 14.3, 3.61, 1.43$, respectively. The upper, middle and lower curves are their corresponding theoretical results, according to Eq.~\ref{eq:final solution}. The analytical results are in agreement with the data, except for the lower curve when $\beta = 1.43$; this is because the PWL model loses accuracy near the diodes' turn-on voltage $U_t$.

\begin{figure}[h]
\centering
\subfigure[]{
 \includegraphics[width = 0.45\textwidth ]{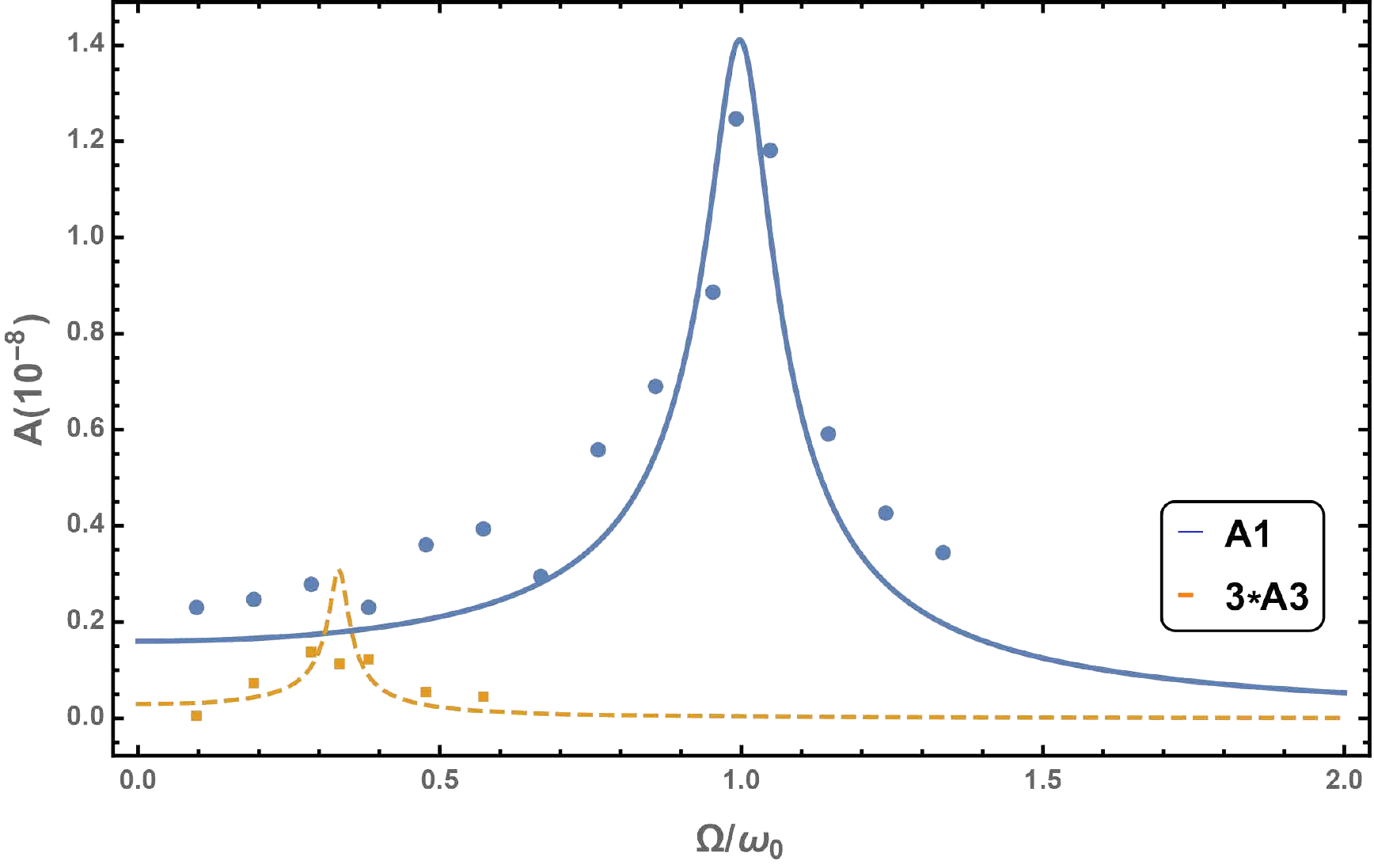}
 \label{fig:fft resonant response}
 }
 \subfigure[]{
 \includegraphics[width = 0.45\textwidth]{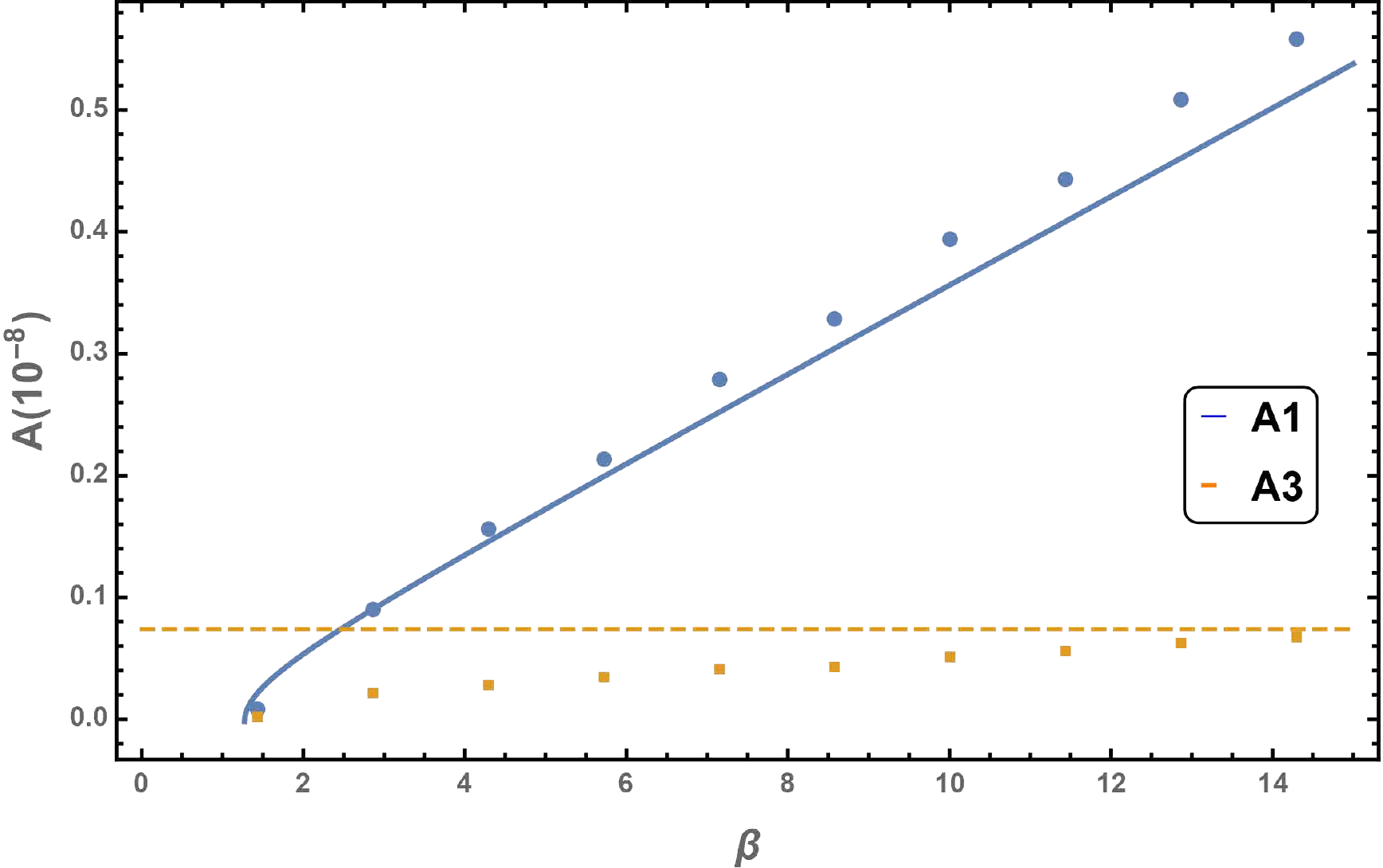}
 \label{fig:amplitude-driving-force}
 }
 \caption{(a)Amplitudes $A_1$ and $A_3$ as a function of driving frequency, $\beta = 7.1$. (b) Amplitudes $A_1$ and harmonic $A_3$ as a function of driving amplitude $\beta$. $\frac{\Omega}{\omega_0} = 0.33$ }
 \label{fig:correlation}
 \end{figure}

In the experiment, the amplitudes of harmonic components are collected using the $FFT$ math function of the digital oscilloscope, shown as the points in  Fig.~\ref{fig:fft resonant response} and Fig.~\ref{fig:amplitude-driving-force}. The curves in Fig.~\ref{fig:fft resonant response} and Fig.~\ref{fig:amplitude-driving-force} show the theoretical resonant amplitudes $A_1$ and harmonic component $A_3$ as a function of driving frequency $\Omega$ and driving amplitude $\beta$, according to Eq.~\ref{eq:final solution}.  The trend derived from Eq.~\ref{eq:final solution} for $A_1$ (see solid curve in  Fig.~\ref{fig:amplitude-driving-force}) is verified by experimental results. In theory, $A_3$ is independent of $\beta$ as indicated by Eq.~\ref{eq:final solution}, but in the experiment, there exists a  weak dependence between  $A_3$ and $\beta$.  Because in the analytical model the static friction $F_0$ (or the turn-on voltage of the diodes $U_t$ in the PWL model) is a constant; however, in the experiment $U_t$ varies with the driving voltage. 



 \subsection{Phase lag and hysteresis}
The phase-frequency response for a linear LC circuit can be illustrated by $U_c$'s  response to the driving voltage $U$ viewed in the X-Y mode of the oscilloscope. When the frequency is much lower/higher than the natural frequency, the phase of $Uc$ and $U$ is nearly the same. The phase lag is $\frac{\pi}{2}$ on resonance.\cite{ref23,feyman} In our nonlinear oscillator model, when the driving amplitude and frequency are sufficiently high, the phase-frequency response is similar to that of a linear resonator, see Figs.~\ref{fig:group imags}(3b)-(3c). Nevertheless, the phase lag is also amplitude dependent. In Figs.~\ref{fig:group imags}(3b)-(3c), when the frequency is lower/higher than the natural frequency, the phase lag decreases/increases with the driving amplitude. Hysteresis occurs when the frequency or the driving amplitude is sufficiently low. As Fig.~\ref{fig:group imags}(3a) shows, when the frequency is low ($\frac{\Omega}{\omega_0} = 0.29$), the $U_c-U$ curve forms hysteresis loop similar to that of ferromagnetic/ferroelectric materials. Sebald et al. also modeled the hysteresis loop of ferroelectric material using dry friction \cite{ref23}. The hysteresis is a result of discontinuity of dry friction when the velocity changes signs. Till now, our solution only considers sliding motion without sticking. Sticking refers to the oscillator remaining still for a short time during its oscillation. Now we investigate the conditions under which the oscillator may slide without sticking, i.e. the stick-slip boundary.

\subsection{Stick-Slip boundary}
If sticking does not happen, once the oscillator reaches a maximum $+x$ position, where $\dot{x} = 0$, then the velocity $\dot{x}$ will change from positive to negative. Then, according to Eq.~\ref{eq:dimensionless eq} the acceleration immediately after  $\dot{x}$ changes signs is 
\begin{equation}
\ddot{x} = -\omega_0^2x + 1 +\beta {\rm cos}(\Omega t_m + \phi_0)
\label{eq:acceleration}
\end{equation}
where $t_m$ is the time when $\dot{x} = 0^-$. In Eq.~\ref{eq:acceleration}, the condition for pure sliding is $\ddot{x} < 0$, which ensures that the oscillator turns back without sticking. If $\ddot{x} > 0$, the sliding friction term "+1" in Eq.~\ref{eq:acceleration} would be unreasonable, since the sliding friction can not drive the oscillator forward. In this case, static friction $f$ should replace the +1 in Eq.~\ref{eq:acceleration}, where  $f$( $|f|<1$) is set so that $\ddot{x} = 0$, and temporary sticking occurs. 
In order to find this stick-slip boundary, we adopt the approximations
that $x \approx A_1 {\rm cos}\Omega t$, and that $t_m \approx \frac{2n\pi}{\Omega}$.
Then the approximate critical condition for stick-slip boundary is
\begin{equation}
-\omega_0^2A_1 + 1 + \beta {\rm cos}\phi_0 = 0
\label{eq:critical condition}
\end{equation}
where $A_1$, $\phi_0$ and $Z$ are defined in Eq.~\ref{eq:final solution}.

\begin{figure}[h!]
    \centering
    \subfigure[]{
    \includegraphics[width = 0.42\textwidth,height = 0.28\textwidth]{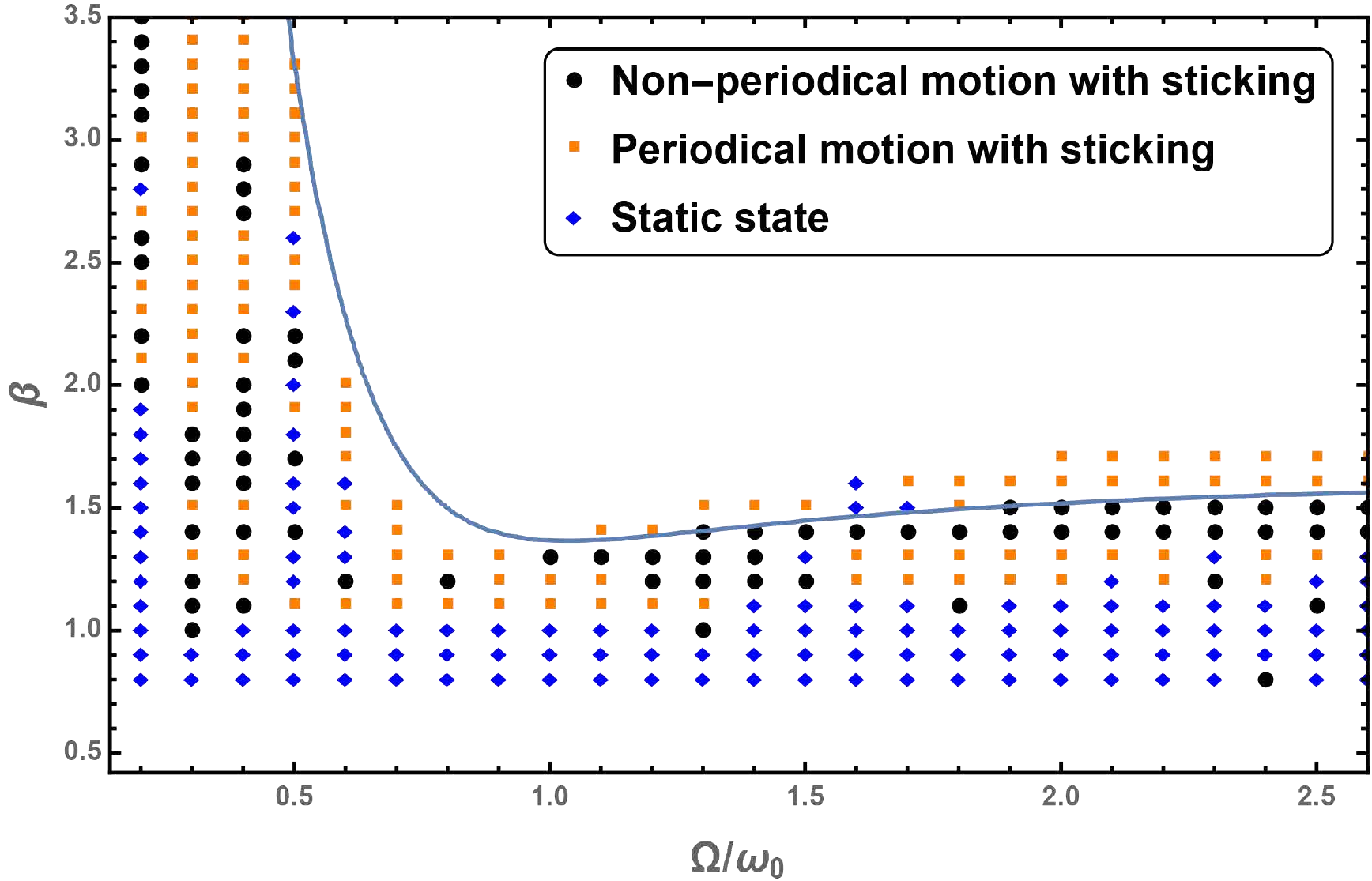}
    \label{fig:boundary Step}
    }
    \subfigure[]{
    \label{fig:circuit_diagram:PWL}
    \includegraphics[width = 0.42\textwidth,height = 0.28\textwidth]{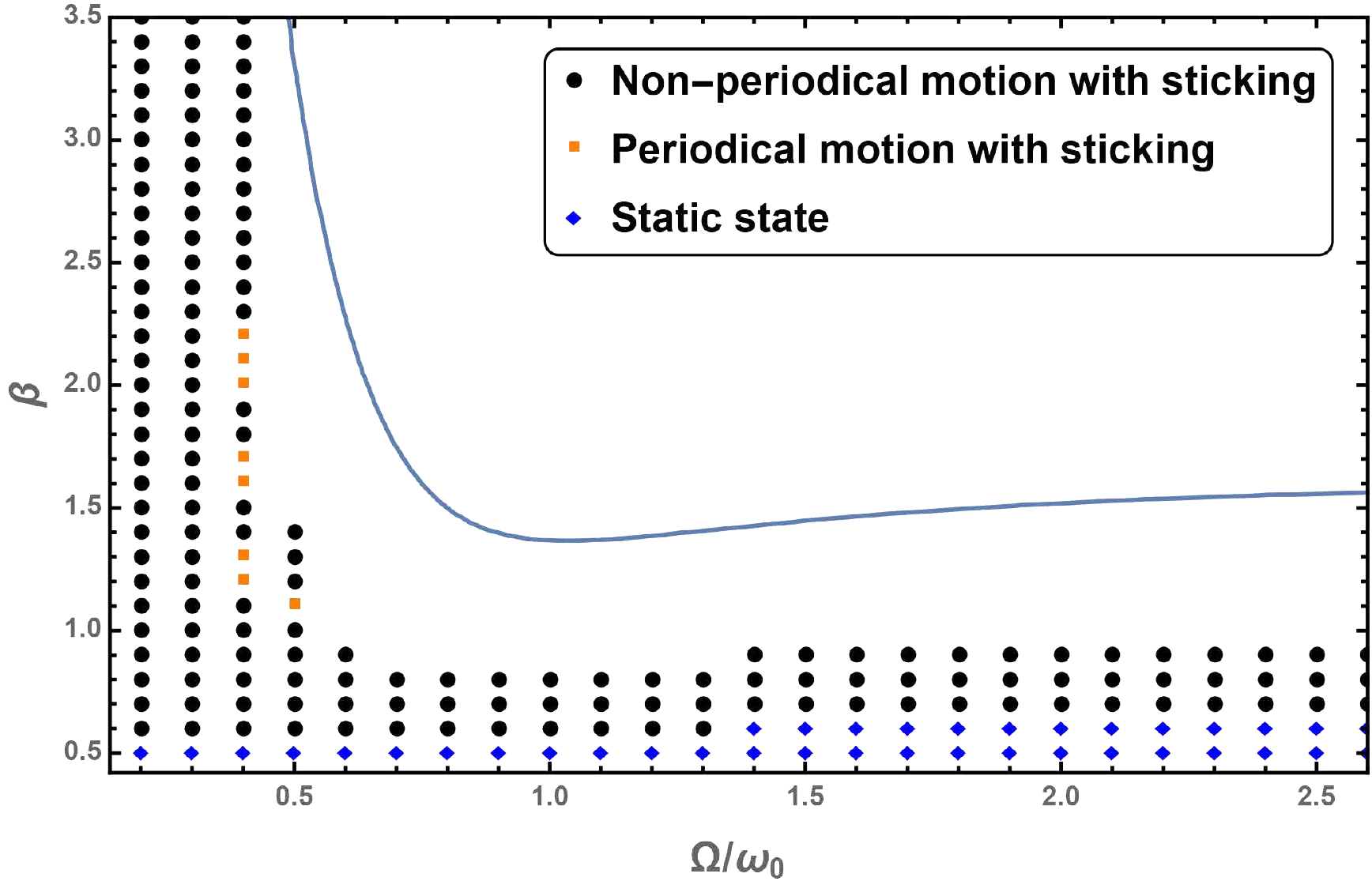}
    \label{fig:boundary Dio}}
    \caption{(a) Stick-slip boundary using PWL model. (b) Stick-slip boundary using the Shockley ideal diode equation. }
    \label{fig:stick-slip boundary}
\end{figure}

The boundaries given by Eq.~\ref{eq:critical condition} are shown as the solid curves in Fig.~\ref{fig:stick-slip boundary}. To find more accurate results, we also make numerical calculations using a LSODA approach. 

\begin{figure}[h!]
     \centering
     \subfigure{
     \includegraphics[width = 0.4\textwidth,height = 0.24\textwidth]{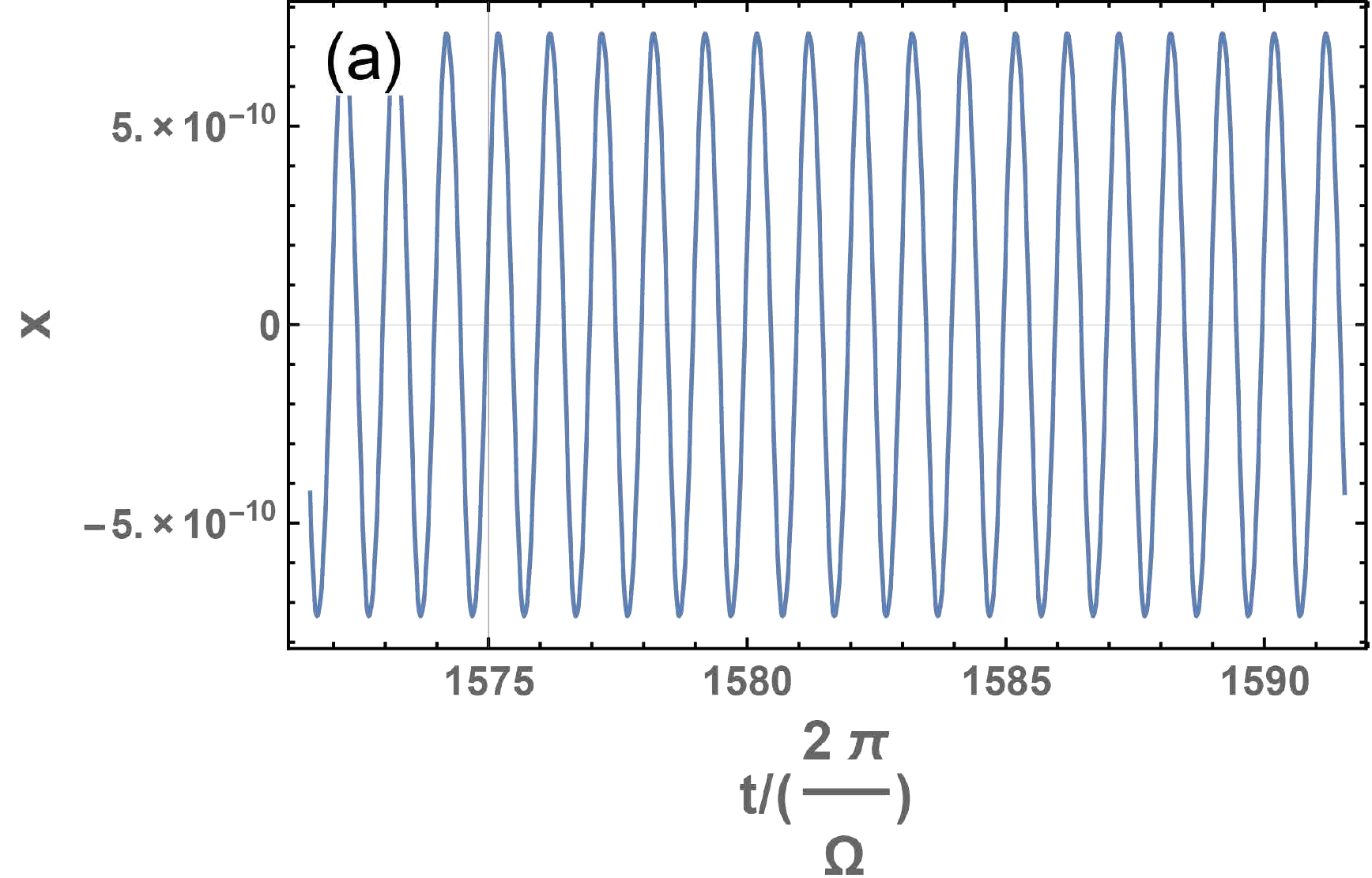}
     \label{fig:periodical}
     }
     \subfigure{
     \includegraphics[width = 0.4\textwidth,height = 0.24\textwidth]{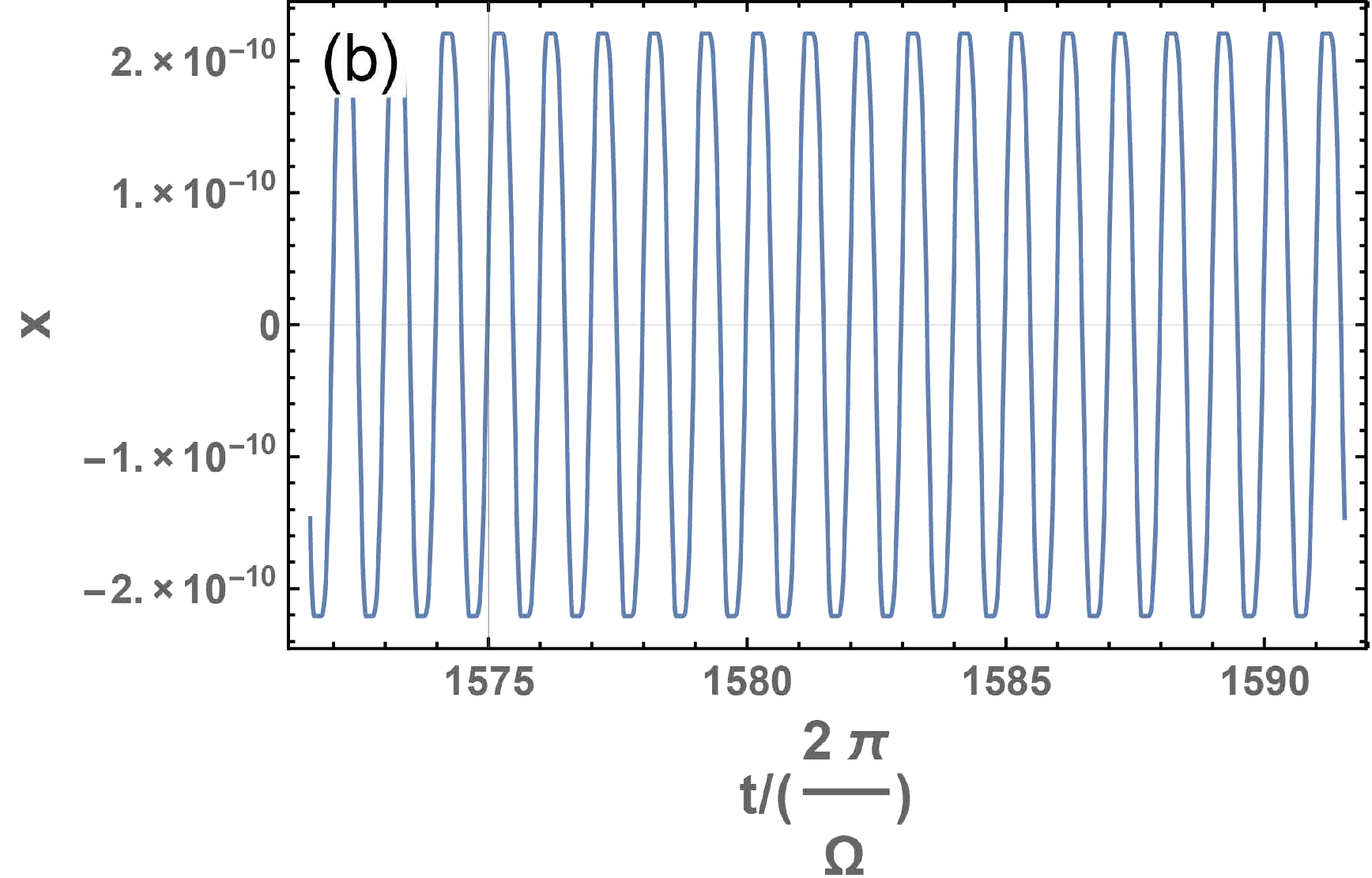}
     \label{fig:nonperiodical}
     }
     \hspace{1in}
     \subfigure{
     \includegraphics[width = 0.4\textwidth,height = 0.24\textwidth]{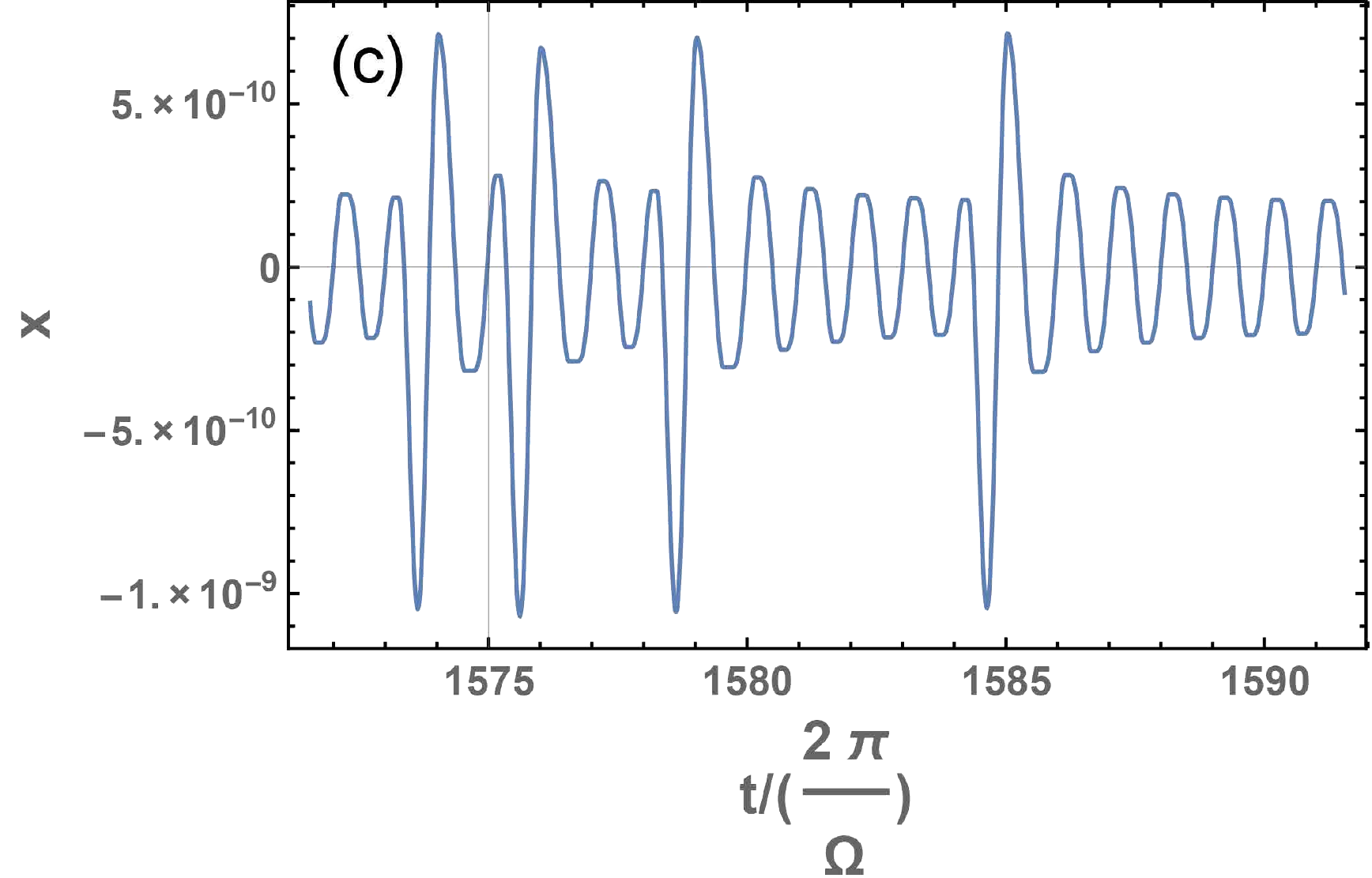}
     \label{fig:sticking}
     }
     \subfigure{
     \includegraphics[width = 0.4\textwidth,height = 0.24\textwidth]{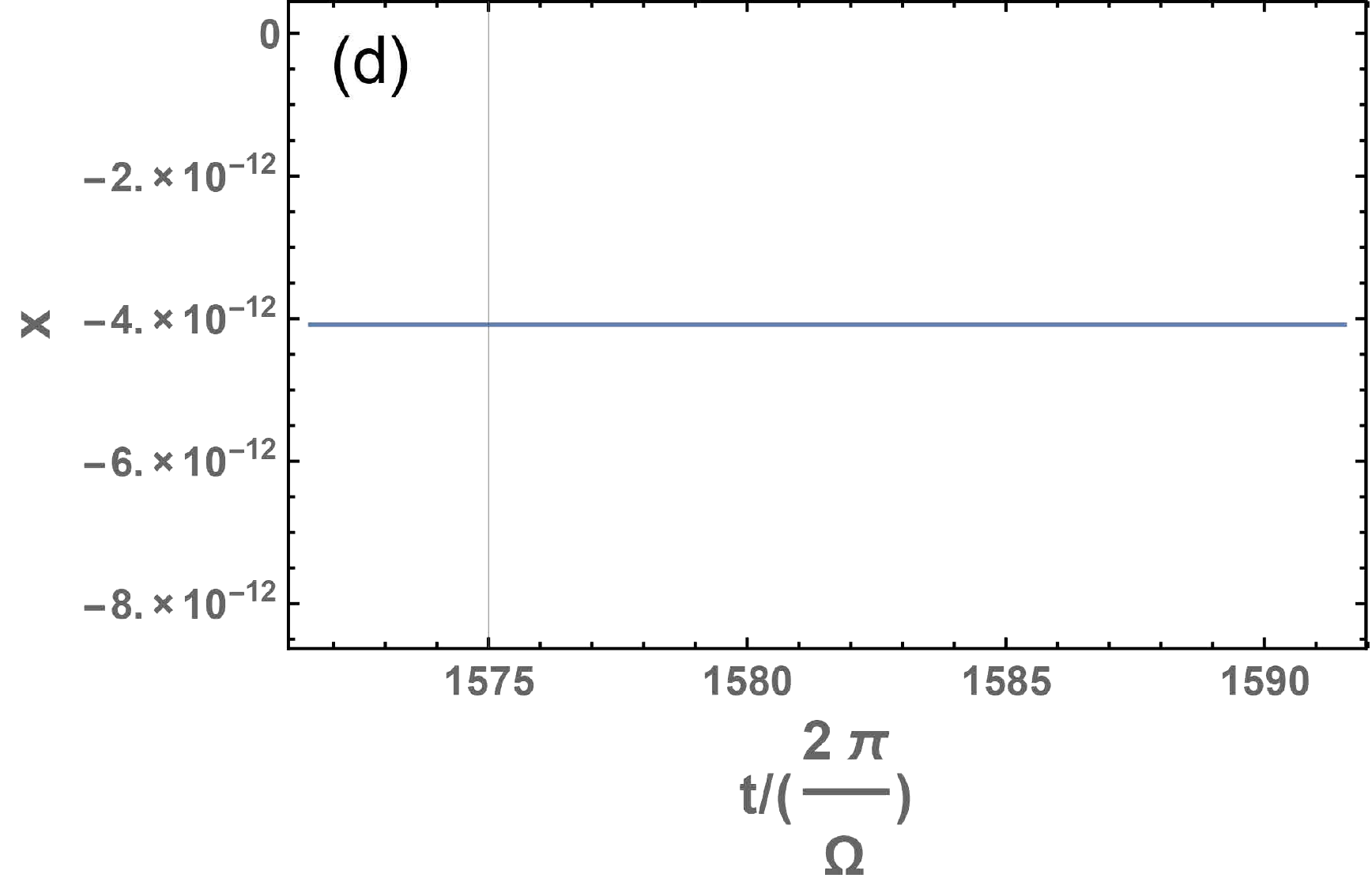}
     \label{fig:still}
     }
     
    \caption{Different simulated oscillation curves. (a)Oscillation curve of the pure sliding motion; $\frac{\Omega}{\omega_0} = 0.9, \beta = 1.6$.
    (b)Oscillation curve of the periodical motion with sticking; $\frac{\Omega}{\omega_0} = 0.6, \beta = 1.2$. (c)Oscillation curve of the non-periodical motion with sticking; $\frac{\Omega}{\omega_0} = 0.7, \beta = 1.2$. 
    (d)Static state; $\frac{\Omega}{\omega_0} = 0.7, \beta = 1.0$.}
    \label{fig:boundary curves}
\end{figure}

There are four types of motion occurring in our simulation whose oscillation curves are shown in Fig.~\ref{fig:boundary curves}: pure sliding motion (Fig.~\ref{fig:boundary curves}(a)), periodical motion with sticking (Fig.~\ref{fig:boundary curves}(b)), non-periodical motion with sticking (Fig.~\ref{fig:boundary curves}(c)) and static state (Fig.~\ref{fig:boundary curves}(d)). The phase diagram of these types of motion and the stick-slip boundary in the parameter space are shown in Fig.~\ref{fig:stick-slip boundary}. The solid line is the theoretical stick-slip boundary given by Eq.~\ref{eq:critical condition} and the dots are simulation results.

Data points in Fig.~\ref{fig:boundary Step} show the calculation results of the oscillator when PWL model is used to characterize the anti-parapllel diodes. When the driving amplitude is large and the driving frequency is near or higher than the natural frequency, the oscillator oscillates without sticking (the blank area). When the driving frequency is much lower or higher than the natural frequency and the driving amplitude is small, the solution is either periodical with zero velocity at turning points for a while (squares) or non-periodical (rounds). These two types of motion are both triggered by the sticking of the oscillator at turning points.\cite{ref24} In the sticking area, numerical solutions are unstable and very sensitive to initial conditions and the oscillator exhibits different kinds of nonlinear behaviors, which are not our focus in this paper. The diamonds represent oscillations with amplitude smaller than hundredth of the driving amplitude $\beta$, thus can be viewed as static.

Fig.~\ref{fig:boundary Dio} shows the calculation results when the Shockley ideal diode equation is used to characterize the anti-parallel diodes. The points represent the same types of motions as in Fig.~\ref{fig:boundary Step}. Here we use the Shockley ideal diode equation\cite{wiki_Scho} to describe the voltage-current characteristic of the diodes,
\begin{equation}
    U = V_{th} {\rm arcsinh}\frac{I}{2I_0}
    \label{eq:Shockley model}
\end{equation}
where $U$ is the voltage across the diodes, $V_{th}$ is the thermal voltage, $I$ is the current through the diodes, and $I_0$ is the reverse bias saturation current. In this part, the term ${\rm sgn}(\dot{x})$ in Eq.~\ref{eq:dimensionless eq} is replaced by ${\rm arcsinh}\frac{\dot{x}U_t}{2I_0 L}$. Compared to the mechanical oscillator whose dry friction is modeled with a sign function, this oscillator is more stable in the sticking area and mostly exhibits periodical motion with a brief standstill at turning points. Besides, the pure sliding region is broader  than that of the oscillator when PWL model is used. 

\section{Conclusion}
We use an LC circuit damped by anti-parallel diodes to simulate the nonlinear behavior of mass-spring systems damped by dry friction. The differential equation for electric oscillator is equivalent to the mechanical one’s when piecewise linear model is used.
We solve the differential equation using series expansand describe the resonant response as well as the amplitudes of superharmonic components. In our experiment, we observe the nonlinear oscillation curves, the frequency spectrum of harmonics, and the hysteresis phenomenon on a digital oscilloscope. We discuss the sticking behaviors of the oscillator and conduct theoretical analysis along with numerical calculations to explore the stick-slip boundary.

\end{document}